\documentclass[10pt,  aps, prd, nofootinbib, twocolumn, noeprints, superscriptaddress]{revtex4-2}

\usepackage[mathlines]{lineno}
\usepackage{amsmath}
\usepackage{amssymb}
\usepackage{bm}      
\usepackage{braket}
\usepackage{dcolumn} 
\usepackage{diagbox} 
\usepackage{epsfig}
\usepackage{graphicx} 
\usepackage{hhline}
\usepackage{mathrsfs}
\usepackage{newtxmath, newtxtext}
\usepackage{slashed}
\usepackage{subfigure}
\usepackage[usenames,dvipsnames]{xcolor}
\usepackage{mathtools, cuted}
\usepackage{lipsum}

\usepackage[pagebackref=false, colorlinks=true]{hyperref}
\usepackage{cleveref}
\definecolor{reddish}{rgb}{0.7,0.2,0.0}  
\definecolor{blueish}{rgb}{0.1,0.1,1}
\hypersetup{
	linkcolor=reddish,      
	citecolor=blueish,      
	filecolor=blue,         
	urlcolor=magenta}       

\begin{document}
	
	\title{Spacetime Measurements with the Photon Ring}
	
	\author{Rahul Kumar Walia}\email{rahulkumar@arizona.edu}
	\affiliation{Department of Physics, University of Arizona, 1118 E 4th Street, 85721 Tucson, Arizona, USA}
	\author{Prashant Kocherlakota}
	\affiliation{Black Hole Initiative at Harvard University, 20 Garden Street, Cambridge, Massachusetts 02138, USA}
	\affiliation{Center for Astrophysics, Harvard \& Smithsonian, 60 Garden Street,
		Cambridge, Massachusetts 02138, USA}
	\author{Dominic O. Chang}
	\affiliation{Department of Physics, Harvard University, Cambridge, Massachusetts 02138, USA}
	\affiliation{Black Hole Initiative at Harvard University, 20 Garden Street, Cambridge, Massachusetts 02138, USA}
	\author{Kiana Salehi}
	\affiliation{Perimeter Institute for Theoretical Physics, 31 Caroline Street North, Waterloo, Ontario, N2L 2Y5, Canada}
	\affiliation{Department of Physics and Astronomy, University of Waterloo, 200 University Avenue West, Waterloo, Ontario, N2L 3G1, Canada}
	\affiliation{Waterloo Centre for Astrophysics, University of Waterloo, Waterloo, Ontario N2L 3G1 Canada}
	
	\begin{abstract}
		We explore the universal symmetries of the black hole photon ring in a wide range of non-Kerr spacetimes, including the Kerr-Newman, Kerr-Sen, Kerr-Bardeen, and Kerr-Hayward metrics. The demagnification exponent ($\gamma$) controls the size and flux scaling of higher-order images, which appear in the photon ring, the time delay ($\tau$) determines the timing of their appearance, and the rotation parameter ($\delta$) relates their relative orientations on the image plane. Our investigation reveals that these critical parameters respond distinctly to variations in black hole spin, generalized charge, and observer inclination, establishing them as complementary probes of spacetime geometry: $\gamma$ is predominantly influenced by charge and spin, $\tau$ is strongly affected by inclination, especially for  near-extremal black holes, and $\delta$ is highly sensitive to spin. Notably, we find that the time delay provides an independent constraint on shadow size for polar observers, while the rotation parameter facilitates metric-independent spin measurements. Specifically, for Kerr black holes, the total variation in $\gamma$, $\tau$, and $\delta$ across all possible inclinations and spins is $\lesssim 20\%$, $\lesssim 10\%$, and $\lesssim 60\%$, respectively. By contrast, the Kerr shadow radius varies by only $\lesssim 8\%$. A future joint measurement of these critical parameters---along with the black hole shadow size---will enable precise spacetime characterization, including measurements of the spin, inclination, and generalized charge.
	\end{abstract}
	
	\maketitle
	
	\section{Introduction} 
	\label{Sec-1}
	
	Black holes (BHs) provide a rare and powerful laboratory to probe both gravitational physics and astrophysics simultaneously. Horizon-scale observations in the electromagnetic spectrum have recently become possible with the Event Horizon Telescope (EHT). These high-resolution, millimeter-wavelength observations of the supermassive BHs in Messier 87 (M87*) and in the Milky Way (Sgr A*) produced the historic first images of BH ``shadows", and are transforming our understanding of their environments and of the curved spacetime near event horizons \cite{EHTC+2019a, EHTC+2022a, EventHorizonTelescope:2024dhe}. 
	The EHT images consistently display a ring-shaped feature known as ``emission ring", whose diameter closely traces the ``critical curve" of the BH shadow boundary \cite{Narayan+2019, Bronzwaer+2021, Kocherlakota+2022, Ozel+2021, Younsi+2021, EHTC+2022f}. 
	Due to limited angular resolution, the EHT has not yet been able to measure the spins of these BHs---a fundamental property with significant implications for BH formation, growth, accretion dynamics, galaxy-scale jet formation and even evolution on galactic scales. This is further complicated by the weak dependence of the size of the BH shadow on spin and inclination, varying by only $\sim 8\%$ for Kerr BHs \cite{Hioki2009,Psaltis+2020}.
	
	General relativity (GR) predicts that the strong gravitational lensing of extended emission sources around BHs produces a direct image as well as higher-order images, with the latter collectively referred to as the ``photon ring" \cite{Bardeen1973, Luminet:1979nyg, Gralla:2019xty, Johnson:2019ljv, Gralla:2019drh}. The photon ring has not yet been resolved in the EHT images. For optically thin emission, the photon ring comprises an infinite sequence of \textit{subrings}---formed by photons that complete at least half of a polar or latitudinal orbit around the BH photon shell before reaching our sky. These subrings display a self-similar structure, and are indexed by the number of polar turning points or half polar-orbits completed en route to the observer, converging exponentially to the critical curve \cite{Gralla:2019drh,Johnson:2019ljv}. Because photons travel along null geodesics of spacetime, these subrings carry distinct imprints of the near-horizon BH geometry, making them a ``sweet spot" for studying the gravitational physics of BHs: They probe the near-horizon region, the leading-order subring remains observable at beyond-Earth baselines despite high redshift, and, importantly, they are minimally affected by the complex astrophysical processes in the accretion disk and jet base.
	
	Photon rings are consistently seen in 3D general relativistic magnetohydrodynamic simulations of hot accretion flows around Kerr BHs \cite{Johnson:2019ljv,Gralla:2020yvo,Paugnat+2022,Cardenas-Avendano:2023dzo,Palumbo:2023auc} as well as around non-Kerr BHs \cite{Chatterjee+2023a, Chatterjee+2023b}. Building on current capabilities, upcoming missions such as the next-generation EHT (ngEHT) \cite{Johnson+2023} and the Black Hole Explorer (BHEX) \cite{Johnson:2024ttr} aim not only to detect the leading-order photon subring in the images of M87* and Sgr A* but also to measure its size and shape (see, e.g., Refs. \cite{Johnson:2024ttr,Galison:2024bop,Kawashima:2024svy, Lupsasca:2024xhq,Akiyama:2024msp}). Detecting the photon ring would confirm a strong prediction of GR, while measuring its geometric features (diameter, width, and shape) and visibility domain features would provide a BH mass and spin measurement, and potentially any deviation from the Kerr geometry, independent of astrophysical assumptions. 
	
	Non-Kerr BHs, both in GR or modified theories of gravity, feature ``deviation parameters'' or  ``generalized charges'' in addition to mass and spin. The observable effects of these deviation parameters are expected to manifest in the strong-field regime near the event horizon. The EHT measurements of the shadow sizes of M87* and Sgr A* have been widely utilized to place constraints on potential deviations from the Kerr metric \cite{EHTC+2019f, EHTC+2022f,Psaltis+2020, Kocherlakota+2021, EHTC+2022f,KumarWalia:2022aop,Vagnozzi:2022moj,Kumar:2020yem,Glampedakis:2023eek}. 
	However, relying on a single observable often results in a degeneracy between the measurements of spin and deviation parameters from the Kerr solution. A Kerr BH shadow of a given size can be very well explained by a non-Kerr BH with different spin and nonzero charge \cite{Kumar:2018ple, Tsukamoto:2014tja}.
	
	The prospect of observing photon ring in the near future has sparked significant interest \cite{Gralla:2019xty,Johnson:2019ljv,Gralla:2019drh,Gralla:2020yvo}. While photon rings for Kerr BHs are relatively well-understood in terms of their image structure, polarization, and Fourier visibility \cite{Gralla:2020yvo, Gralla:2020nwp,Vincent+2022,Kocherlakota+2024a,Kocherlakota+2024b,Gralla:2019drh,Gralla:2019drh,Johnson:2019ljv,Paugnat+2022,Cardenas-Avendano:2023dzo,Hadar+2021,Hadar+2022,Himwich:2020msm,Jia:2024mlb}, their morphology in non-Kerr spacetimes is not fully understood yet (see, however, Refs. \cite{Staelens:2023jgr,Kocherlakota+2024a,Kocherlakota+2024b,Wielgus:2021peu,daSilva:2023jxa, Cunha:2017qtt, Medeiros:2019cde}). In this paper, we aim to advance our current understanding of photon rings by addressing some key questions: How does the structure of the photon ring vary across different BH spacetimes? Can detections of the photon ring lead to robust BH spin measurement? And, most importantly, can joint measurements of the shadow size and specific photon ring features provide stringent constraints on deviations from Kerr spacetime?
	
	Building on Ref.~\cite{Gralla:2019drh}, in our previous paper \cite{Salehi:2024cim}, we have described the three critical parameters ($\gamma_{\rm p}, \tau_{\rm p}, \delta_{\rm p}$) that govern the demagnification, time delay, and rotation, respectively, of successive photon subrings on the image plane in a broad class of stationary and axisymmetric spacetimes. In Ref.~\cite{Salehi:2024cim}, we employed the Johannsen metric~\cite{Johannsen:2013szh}---a four-dimensional, geodesically integrable, spinning metric---to determine the critical parameters in terms of two metric functions and the observer's inclination angle.
	\footnote{The Johannsen metric \cite{Johannsen:2013szh} is described by four independent metric functions $f, F, N, B$ in general. However, we have shown in Sec. II of Ref. \cite{Salehi+2023} that the metric function $f$ can be absorbed into the ``Mino time'' and that the metric function $B$ can be eliminated via a radial coordinate transformation. Thus, null geodesics in the Johannsen metric and the critical lensing parameters are completely determined by $F$ and $N$.} %
	This metric is widely adopted for studying strong-field effects in a theory-agnostic framework, as it can represent a broad range of non-Kerr BHs in GR and modified theories. This paper explores the astrophysical implications of these critical parameters. We first systematically analyze their variation with the spin and inclination of Kerr BHs and then quantitatively illustrate how deviations from the Kerr solution affect these parameters. To systematically investigate variations in photon ring morphology across different spinning spacetimes, we analyze specific BH solutions, including Kerr-Newman \cite{Newman:1965my}, Kerr-Sen \cite{Sen:1992ua}, Kerr-Bardeen \cite{Bambi:2013ufa}, and Kerr-Hayward \cite{Bambi:2013ufa}. These non-Kerr spacetimes are controlled by an additional ``charge'' parameter $Q$, arising from their underlying physical fields, in addition to the total mass $M$ and spin $a$. All of these metrics reduce to the Kerr metric in the limit of vanishing charge. Through this analysis, we quantitatively assess the potential of photon ring measurements to constrain strong-field deviations from the Kerr geometry, with the following key outcomes:
	\begin{enumerate}
		\item The demagnification exponent $\gamma_{\rm p}$ is always smaller than its Schwarzschild BH value ($\pi$), implying that photon subrings in non-Kerr spacetimes are always wider and brighter than in a Schwarzschild BH spacetime.
		\item The time delay $\tau_{\rm p}$ provides an independent measurement of the BH shadow size for observers present close to its spin-axis (e.g., M87$^*$).
		\item The rotation parameter $\delta_{\rm p}$ yields a robust metric-independent spin measurement.
	\end{enumerate}
	Thus, higher-order images, i.e., photon subrings, serve as ``fingerprints" of the spacetime geometry, as they are governed by three purely geometric critical parameters.

	The paper is organized as follows: In Section~\ref{Sec-2}, we present a brief review of the definitions of the lensing critical parameters in this broad class of stationary and axisymmetric, integrable BH spacetimes. In Section~\ref{Sec-3}, we explore the variation of these critical parameters in various non-Kerr spacetimes, examining both the impact of deviation parameters and observer's inclination angle. We discuss the implications for BH spin measurements and constraints on Kerr-deviation parameters that could be obtained from future photon ring observations. Finally, we summarize our findings in Section~\ref{Sec-4}.
	
	\section{Photon Ring Critical Parameters}\label{Sec-2}
	
	The Johannsen metric in Boyer-Lindquist coordinates, $x^\mu = (t, r, \vartheta, \varphi)$, is expressed  as \cite{Johannsen:2013szh,Salehi+2023}
	\begin{align} \label{eq:JP_Metric}
		\mathrm{d}s^2 =&\ 
		- \frac{\Sigma}{\Pi^2}(N^2 - F^2 a^2\sin^2{\vartheta})\mathrm{d}t^2 \\
		&\ - 2\frac{\Sigma}{\Pi^2}(rF-N^2)a\sin^2{\vartheta}~\mathrm{d}t\mathrm{d}\varphi 
		\nonumber \\
		&\ + \frac{\Sigma}{\Pi^2}(r^2 - N^2 a^2\sin^2{\vartheta})\sin^2{\vartheta}~\mathrm{d}\varphi^2 \nonumber \\
		&\ + \frac{\Sigma B^2}{r^2 N^2}\mathrm{d}r^2 + \Sigma\mathrm{d}\vartheta^2\,, \nonumber 
	\end{align}
	where
	\begin{align}
		&\Sigma(r, \vartheta) = r^2 + f(r) + a^2\cos^2{\vartheta},\\
		&\Pi(r, \vartheta) = r - F a^2\sin^2{\vartheta}.
	\end{align}
	The metric functions $N, F$, and $f$ can be freely chosen to describe a large class of integrable spacetimes, not restricted to only BH spacetimes. The metric function $B$ can be eliminated through a change of the radial coordinate, $\mathrm{d}\rho = B(r)\mathrm{d}r$. 
	
	The coordinate $4-$velocity of an arbitrary photon, $k^\mu = \mathrm{d}x^\mu/\mathrm{d}\lambda_{\rm m}$, where $\lambda_{\mathrm{m}}$ is the Mino time along its trajectory, is 
	\begin{equation} \label{eq:NG_Tangent}
		k^\mu = (\mathscr{T}_r + a^2\cos^2{\vartheta},\, \pm_r\sqrt{\mathscr{R}}, \,\pm_\vartheta\sqrt{\Theta},\, \Phi_r + \xi\csc^2{\vartheta})\,.
	\end{equation}
	In the above, we have introduced the so-called null geodesic effective potentials%
	\footnote{This is inspired by equations of the form $\dot{r}^2 = \mathscr{R}$ or $\dot{\vartheta}^2 = \Theta$, which are akin to ``energy equations.''} %
	$\{\mathscr{T}_r(r), \mathscr{R}(r), \Theta(\vartheta), \Phi_r(r)\}$,
	\begin{align} \label{eq:Eff_Pots}
		\mathscr{T}_r =&\ \frac{r}{N^2}(r - Fa\xi) + a\xi-a^2\,,
		\\
		\mathscr{R} =&\ \frac{r^2}{B^2}\left[(r - Fa\xi)^2 - N^2\mathscr{I}^2\right]\,, 
		\nonumber \\
		\Theta =&\ \mathscr{I}^2 - (\xi\csc{\vartheta} - a\sin{\vartheta})^2\,,
		\nonumber \\
		\Phi_{r} =&\ \frac{aF}{N^2}(r - F a \xi) - a\,,
		\nonumber
	\end{align}
	with $E$ the conserved energy of the photon, $E\xi = L$ its conserved azimuthal angular momentum, and $\mathscr{I}$ its dimensionless non-negative Carter constant. Adding a constant to the Carter constant yields another Carter constant, e.g., 
	\begin{equation}
		\eta = \mathscr{I}^2 - (\xi - a)^2\,.
	\end{equation}
	Note, however, that shifting the Carter constant in this way does not change the photon orbit. Equations~(\ref{eq:NG_Tangent}) and (\ref{eq:Eff_Pots}) reveal that knowledge of the metric functions $N$ and $F$ is sufficient to determine the orbits of arbitrary photons in a Johannsen spacetime (we can set $B=1$ through a change of $r$, as discussed above).
	
	The conserved quantities $(\xi, \mathscr{I})$ can be used to understand image formation. If we set up Cartesian coordinates $(\alpha, \beta)$ to describe the image plane of a faraway observer located at an inclination $\vartheta = \mathscr{i}$ relative to the BH spin axis ($\vartheta = 0$), then the location at which an arbitrary photon appears is given simply as \cite{Bardeen1973},
	\begin{equation} 
		\alpha = -\xi\csc{\mathscr{i}}\,;\ \ 
		\beta = \pm_\vartheta\sqrt{\Theta(\mathscr{i})}\,,
	\end{equation}
	where the sign $\pm_\vartheta$ corresponds to sign of the polar angular momentum $k_\vartheta$ at the location of the observer. Thus, the conserved quantities $(\xi, \mathscr{I})$ directly determine the impact parameters of the photon. BH shadow boundary curve or the critical curve for parameterized metric (\ref{eq:JP_Metric}) has been extensively studied in literature \cite{Younsi:2016azx, Konoplya:2021slg, Younsi:2021dxe, Johannsen:2013vgc}. Furthermore, the chaotic behavior of null geodesics and the resulting BH shadows are explored in the context of nonintegrable spacetimes, revealing complex and intriguing gravitational dynamics \cite{Glampedakis:2018blj,Cunha:2016bjh,Shipley:2016omi,Kostaros:2021usv,Kostaros:2024vbn}.
	
	Strong gravitational lensing close to a BH can compel photons to move on a fixed BL radius orbit---bound spherical orbits. The radial velocity, $|\dot{r}| = \mathscr{R}$, and acceleration, $|\ddot{r}| = \partial_r\mathscr{R}/2$, of photons on such spherical null geodesics (SNGs) both must naturally vanish. This is sufficient to infer the impact parameters $(\xi_{\mathrm{p}}, \mathscr{I}_{\mathrm{p}})$ of an SNG on a sphere of radius $r=r_{\mathrm{p}}$ as being
	\begin{align} \label{eq:SNG_Impact_Parameters}
		\xi_{\mathrm{p}} = \left.\frac{1}{a}\frac{N - r\partial_r N}{N\partial_rF - F \partial_r N}\right|_{r_{\rm p}};\ \ 
		\mathscr{I}_{\mathrm{p}} = -\left.\frac{F - r\partial_r F}{N\partial_rF- F \partial_r N}\right|_{r_{\rm p}}\,.
	\end{align}
	The shape and, in particular, the radial extent of the photon shell is determined simply by demanding that the $4-$velocities of the SNGs are real-valued. For this consideration, the polar angular velocity, $\dot{\vartheta}$, of SNGs alone plays a nontrivial role. Requiring that
	\begin{equation} \label{eq:Th_Nonnegativity}
		\Theta(\vartheta) \geq 0,    
	\end{equation}
	determines the shape of the photon shell in a BH spacetime. 
	
	The equality in Eq.~(\ref{eq:Th_Nonnegativity}) locates the outermost $r=r_{\mathrm{p}}^+$ and innermost $r=r_{\mathrm{p}}^-$ bound null orbits in the BH exterior. These are the retrograde ($\xi_{\mathrm{p}} < 0$) and the prograde ($\xi_{\mathrm{p}} > a$) equatorial circular photon orbits respectively. These radii $r_{\mathrm{p}}^{\pm}$ respectively are the largest roots of the two equations
	\begin{equation} \label{eq:NG_r_Eq}
		\mp a(F - r\partial_r F) + (N - r\partial_r N) - a^2(N\partial_rF - F \partial_r N) = 0\,.
	\end{equation}
	Within the photon shell, there is only one SNG of radius $r=r_{\mathrm{p}}^0$ that reaches the poles and is a zero angular momentum orbit, $\xi_{\mathrm{p}}^0 = 0$. The outer photon shell $r_{\mathrm{p}}^0 < r \leq r_{\mathrm{p}}^+$ permits only retrograde SNGs whereas its inner region $r_{\mathrm{p}}^- \leq r < r_{\mathrm{p}}^0$ permits only prograde ones.
	
	The polar motion of an arbitrary SNG corresponds to oscillations between its two polar turning points $\vartheta = \vartheta_{\pm}$, which are, once again, obtained from the equality in Eq.~(\ref{eq:Th_Nonnegativity}), i.e., $\Theta(\vartheta_{\pm}) = 0$. In terms of $u=\cos^2{\vartheta}$, this equation reduces to a simple quadratic equation in $u$ \cite{Gralla:2019drh},
	\begin{equation} \label{eq:Polar_TPs}
		(1-u)\Theta(u) = \eta - a^2u^2 + (a^2 - \eta - \xi^2)u = 0\,,
	\end{equation}
	the solutions to which are given as
	\begin{equation} \label{eq:u_sol}
		u_\pm = \Delta_\vartheta \pm \sqrt{\Delta_\vartheta^2 + \frac{\eta}{a^2}}\,;\quad 
		\Delta_\vartheta = \frac{1}{2}\left[1 - \frac{\eta+\xi^2}{a^2}\right].
	\end{equation}
	
	Since SNGs have non-negative $\eta$ ($=\Theta(\pi/2)$), we can see that only $u_+ > 0$ for such orbits. Their polar turning points $\vartheta_\pm$ are then given as,
	\begin{equation}\label{eq:thetaTurning}
		\vartheta_\pm = \arccos{\left(\mp\sqrt{u_+}\right)}\,.
	\end{equation}
	
	In addition to the solutions in Eq.~(\ref{eq:u_sol}), the polar turning point in Eq.~(\ref{eq:Polar_TPs}) also has a trivial solution $u=1$, satisfied by zero angular momentum orbits (ZAMOs, $\xi_{\rm p}^0 = 0$). ZAMOs have trivial polar turning points corresponding to the poles of the coordinate system, $\vartheta=0, \pi$. The ZAMO SNG, in particular, is located, as noted above, at $r=r_{\mathrm{p}}^0$.
	
	With the above, we are now in a position to introduce the lensing critical parameters. These are given as \cite{Salehi:2024cim}
	\begin{equation}\label{Eq:Critical-parameter}
		\begin{aligned}
			\gamma_{\mathrm{p}} &\:= \sqrt{\frac{\partial_r^2\mathscr{R}_{\mathrm{p}}}{2}}\hat{G}_\vartheta\,, 
			\\
			\tau_{\mathrm{p}} &\:= \left[\frac{r_{\mathrm{p}}}{N_{\mathrm{p}}}\mathscr{I}_{\mathrm{p}} + a\xi_{\mathrm{p}} - a^2\right]\hat{G}_\vartheta + a^2\hat{G}_t\,, \\
			\delta_{\mathrm{p}} &\:= a\left[\frac{F_{\mathrm{p}}}{N_{\mathrm{p}}}\mathscr{I}_{\mathrm{p}} - 1\right]\hat{G}_\vartheta + \xi_{\mathrm{p}}\hat{G}_\varphi + 2 \pi H(r-r_{\rm p}^0)\,,
		\end{aligned}
	\end{equation}
	where, in the above, $H(x)$ is the Heaviside step function. These critical parameters not only govern the nearly bound photon orbits around the BH but also determine the scaling of higher-order subrings on the image screen. We have also introduced the hatted functions $\hat{G}$ which are given in terms of complete elliptic functions as follows \cite{Kapec+2020, Gralla:2019drh}
	\begin{align}
		\begin{alignedat}{2}
			\hat{G}_\vartheta 
			&:= \int_{\vartheta_-}^{\vartheta_+}\frac{1}{\sqrt{\Theta(\theta)}}\mathrm{d}\vartheta
			&&= \frac{2}{a\sqrt{-u_-}}K\left(\frac{u_+}{u_-}\right)\,,
			\\
			\hat{G}_t 
			&:= \int_{\vartheta_-}^{\vartheta_+}\frac{\cos^2\vartheta}{\sqrt{\Theta(\theta)}}\mathrm{d}\vartheta 
			&&= -\frac{4u_+}{a\sqrt{-u_-}}E'\left(\frac{u_+}{u_-}\right)\,,
			\\
			\hat{G}_\varphi 
			&:= \int_{\vartheta_-}^{\vartheta_+}\frac{\csc^2\vartheta}{\sqrt{\Theta(\theta)}}\mathrm{d}\vartheta 
			&&= \frac{2}{a\sqrt{-u_-}}\Pi\left(u_+ \middle| \frac{u_+}{u_-}\right)\,,
		\end{alignedat}
	\end{align}
	with $K(k)$ the complete elliptic function of the first kind, $E'(k)$ the derivative of the complete elliptic function of the second kind with respect to the elliptic modulus $k$, i.e., $E'(k) := [E(k) - K(k)]/(2k)$, and $\Pi(n|k)$ the complete elliptic integral of the third kind, with $n$ denoting the elliptic characteristic. Our convention for the elliptic functions can be found in Ref. \cite{Salehi:2024cim}.
	
	The prefactor of the term involving $\hat{G}_\vartheta$ in $\gamma_{\mathrm{p}}$ is the (Mino time) phase space Lyapunov exponent, $\kappa_{\mathrm{p}} = (\partial_r^2 \mathscr{R}_{\mathrm{p}}/2)^{1/2}$. This Lyapunov exponent measures the radial instability of the SNGs, and governs the radial deviation, $\delta r(\lambda_{\mathrm{m}})$, of nearly bound photon orbits from their bound counterparts (i.e., those having the same impact parameters),
	\begin{align} \label{eq:Radial_Dev_Sol_Mino}
		\log{\left[\frac{\delta r(\lambda_{\mathrm{m}})}{\delta r(0))}\right]} \approx \pm\kappa_{\mathrm{p}}\lambda_{\rm m}\quad 
		\mathrm{or}\quad
		\delta r(\lambda_{\mathrm{m}})=\delta r(0)\,e^{\pm\kappa_{\mathrm{p}}\lambda_{\rm m}}.
	\end{align}
	In the above, $\delta r(0)$ represents the initial radial coordinate distance between the two orbits. For photons on radially perturbed orbits that are moving away from (toward) the SNG, one picks the positive (negative) sign in the equations above.
	
	The prefactors in the terms involving $\hat{G}_\vartheta$ in $\tau_{\mathrm{p}}$ and $\delta_{\mathrm{p}}$ are simply the values of the effective potentials $\mathscr{T}_r$ and $\Phi_r$ \eqref{eq:Eff_Pots} evaluated at $r=r_{\mathrm{p}}$ respectively.
	
	We have thus far described the critical parameters associated with an arbitrary SNG in the photon shell. The observer's inclination angle, $\mathscr{i}$ (i.e., the observer is located at $\vartheta = \mathscr{i}$), constrains the portion of the photon shell, $\Theta(\mathscr{i})\geq 0$, that contributes to image formation. Essentially, photons with turning points whose interval does not include the observer's colatitude, i.e., $\mathscr{i} \notin [\vartheta_-, \vartheta_+]$ can never reach the observer.
	
	Furthermore, the nearly bound null geodesics that do appear on the image plane appear at different image plane polar angles, $\psi$, where
	\begin{equation}
		\psi = \arctan(\beta/\alpha)\,,\label{Eq:rp-psi}
	\end{equation}
	depending on their impact parameters $(\xi_{\mathrm{p}}, \mathscr{I}_{\mathrm{p}})$. Thus, each of the lensing critical parameters is a function of $\psi$ on the image plane of the observer.
	
	In the following section, we will be interested in understanding the variation of the critical parameters with the observer's inclination. For this purpose, it will be useful to introduce the mean values of the critical parameters over the image plane polar angle as,
	\begin{equation}
		\begin{aligned} \label{eqn:ave_critical_exponents}
			\langle \gamma \rangle_{\psi} &= \frac{1}{2\pi}\int_0^{2\pi} \gamma_{\mathrm{p}}(\psi)\mathrm{d}\psi\,, 
			\\
			\langle \tau \rangle_{\psi} &= \frac{1}{2\pi}\int_0^{2\pi} \tau_{\mathrm{p}}(\psi)\mathrm{d}\psi\,, \\
			\langle \delta \rangle_{\psi} &= \left[\frac{1}{2\pi}\int_0^{2\pi} \delta_{\mathrm{p}}(\psi)\mathrm{d}\psi\right]\ \mathrm{mod}\ 2\pi\,.
		\end{aligned}
	\end{equation}
	It is worth noting here that the rotation parameter $\delta_{\mathrm{p}}$ measures the image screen polar angle offsets between consecutive order images \cite{Gralla:2019drh}. Therefore, it satisfies the following identity $\delta_{\mathrm{p}} = \delta_{\mathrm{p}}\ \mathrm{mod}\ 2\pi$. For this reason, it can be useful to introduce a ``wrapped'' rotation parameter $\delta_{2\pi; \mathrm{p}}$ as 
	\begin{equation}
		\delta_{2\pi; \mathrm{p}} = \delta_{\mathrm{p}}\ \mathrm{mod}\ 2\pi\,.
	\end{equation}
	
	Below, however, we elect to \textit{not} use the wrapped rotation parameter since when averaged, $\langle \delta_{2\pi; \mathrm{p}}\rangle_\psi := (1/2\pi)\int_0^{2\pi} \delta_{2\pi; \mathrm{p}}(\psi)\mathrm{d}\psi$, it introduces unnecessary wrapping artifacts (see right column in Fig.~\ref{fig:critical_exponent_mean_and_median}). Instead, $\langle \delta\rangle_\psi$ defined in Eq.~(\ref{eqn:ave_critical_exponents}) above will describe the mean rotation parameter. The appendix \ref{app:AppA_Measures} discusses this issue in further detail. We also address there the impact on our conclusions of our choice to use the mean as a characteristic measure of the critical parameters on the image plane versus, e.g., the median. While the mean and median measurements of $\delta_{\rm p}$ agree at small inclinations, they differ significantly at large inclinations, where, unlike the mean $\delta_{\rm p}$, the median $\delta_{\rm p}$ remains nearly independent of inclination. This behavior can be partially attributed to the following reasoning: The portion of the photon shell contributing to the photon ring formation grows smoothly with the observer's inclination---ranging from a single SNG, $r_{\rm p}^0$, for a polar observer to the entire photon shell, $r_{\mathrm{p}}^- \leq r < r_{\mathrm{p}}^+$, for an equatorial observer. Thus, at large inclinations, different segments of the photon ring, constructed from distinct SNGs with varying BL radii, experience differential demagnification, delay, and rotation. While the mean, as a statistical average, is sensitive to outliers (extremely high or low values) caused by prograde and retrograde SNGs near the equatorial plane, the median is not. Therefore, the mean  $\delta_{\rm p}$ provides a more reliable statistical average than the median.

	We further note that phase wrapping information is not readily available from snapshots of order$-n$ and order$-(n+1)$ images.
	It is, however, possible to experimentally observe phase wrapping from \textit{movies} of successive order images, by studying how the $n+1$ image changes under continuous motion of the $n$ image. As an example, the phase wrapping associated with an orbiting hot spot could be found by marking images by their time stamps whenever they cross over the same screen angle $\psi$, and keeping note of whether they cross over in a clockwise or anticlockwise motion as the hot spot goes around one revolution. We refer the reader to Fig.~\ref{fig:Kerr_KerrNewman_critical_exponent_variation} to illustrate how phase wrapping information affects $\delta_{\rm p}$.
	
	Since we are simultaneously interested in understanding the variation of the critical parameters with spacetime geometry, we find it useful to work with the mean deviations of the critical parameters from their Schwarzschild BH values, for which they are constant along the photon ring and are expressed as \cite{Gralla:2019drh},
	\begin{equation} \label{eq:Schw_Crit_Parameters}
		\{ \gamma,  \tau, \delta \} = \{\pi, \sqrt{27}\pi M, \pi\}\,,  \end{equation}
	where $M$ denotes the Arnowitt-Deser-Misner (ADM) mass of the spacetime. Throughout this work we assume that the ADM masses of all BH spacetimes considered here are precisely identical. In practice, this mass for both M87$^*$ and Sgr A$^*$ can be fixed via stellar-dynamics measurements \cite{Gebhardt+2011, Gravity+2018}. For BHs having the same ADM mass, we define the mean critical parameter deviations as
	\begin{equation}\label{eqn:deviation_critical_exponents}
		\langle \bar{\gamma} \rangle_{\psi} := \frac{\langle \gamma \rangle_{\psi}}{\pi} - 1\,;\ 
		\langle \bar{\tau} \rangle_{\psi} := \frac{\langle \tau \rangle_{\psi}}{\sqrt{27}\pi M} - 1\,;\ 
		\langle \bar{\delta} \rangle_{\psi} := \frac{\langle \delta \rangle_{\psi}}{\pi} - 1\,.
	\end{equation} 
	
	\section{Variation of Critical Parameters with Observer Inclination and Non-Kerr Spacetimes}\label{Sec-3}
	In this section, we study the behavior of the critical parameters with varying spin, charge and observer inclination for a number of spinning BHs. 
	
	Though our calculations are specific to the limit of high image-order (large$-n$), we note that these results are indicative of measurements that will become accessible from upcoming experiments such as the ngEHT and BHEX \cite{Johnson:2024ttr}, which will resolve the $n=1$ images of M87$^*$ and Sgr A$^*$. Ref.~\cite{Kocherlakota+2024a} has recently analyzed the variation in the exact time delay between the appearance of the primary ($n=0$) and secondary ($n=1$) images of a point source in a Schwarzschild BH spacetime. While the theoretically predicted ``critical" value of time delay in this spacetime is a constant ($\tau = \sqrt{27}\pi M$), the exact time delay depends linearly both on the radial location of the source as well as the relative inclination between source and observer. Nevertheless, for sources located roughly in the equatorial plane viewed from moderate inclinations, the difference between the exact time delay and the critical time delay is modest ($\lesssim 30\%$). We expect this trend to extend to the other critical parameters and to also be relevant to the spinning BH spacetimes considered in this work. Specifically, we anticipate that our large$-n$ results will approximate the exact demagnification, time delay and rotation between the primary and secondary images of hotspots and accretion disks in images of M87$^*$ and likely also Sgr A$^*$. 
	
	\begin{figure*}
		\centering
		\includegraphics[width=\linewidth]{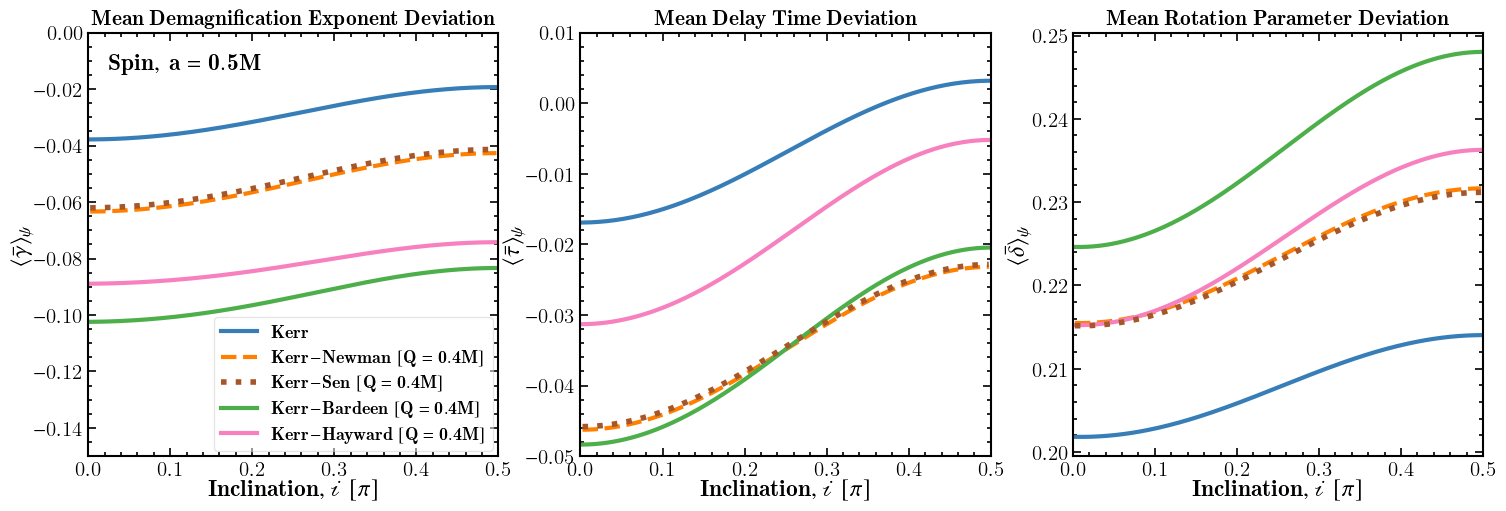}
		\caption{Fractional deviations of the mean demagnification exponent (left panel), time delay (center panel), and rotation parameter (right panel) of the Kerr, Kerr-Newman, Kerr-Sen, Kerr-Bardeen and Kerr-Hayward BHs from their corresponding values for Schwarzschild BH, for varying observer inclinations. The spin, $a$, of each BH is fixed to $0.5 M$ and the ``generalized charge,'' $Q$, of the non-Kerr BHs is set to $0.4 M$. We find modest variations ($\lesssim 5\%$) in all three critical parameters across the different spacetimes.  }
		\label{fig:half spin and extra charge}
	\end{figure*}

	\begin{figure*}
		\centering   
		\includegraphics[width=\linewidth]{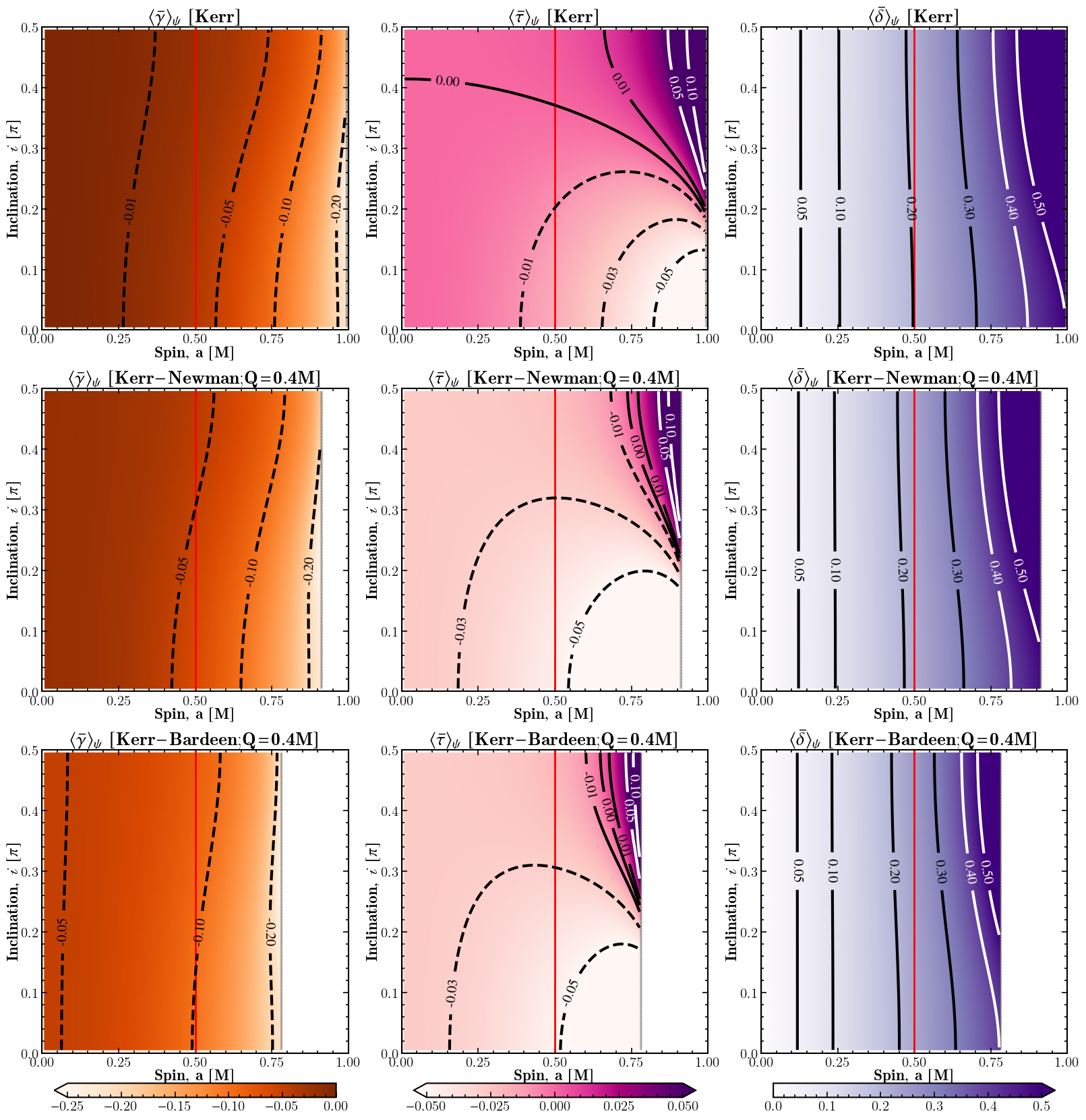}
		\caption{Fractional deviation of the (mean) critical parameters of the Kerr (top row), Kerr-Newman (middle row) and Kerr-Bardeen (bottom row) BHs from their corresponding values for the  Schwarzschild BH, as a function of BH spin and observer inclination.
			The charge of these non-Kerr BHs is set to a middling value of $0.4 M$, as in Fig.~\ref{fig:half spin and extra charge}. Isocontours of critical parameters with constant values smaller (or larger) than the corresponding values for the Schwarzschild BH are shown by dashed (or solid) lines. The demagnification exponent (left column) is always smaller than its Schwarzschild value; thus, secondary images are larger in non-Schwarzschild spacetimes. For most values of ($a, \mathscr{i}$), the time delay (middle column) is smaller than the Schwarzschild BH value but increases sharply for near-extremal BHs viewed by an equatorial observer, due to the influence of photons orbiting extremely close to the event horizon. The rotation parameter (right column) always exceeds the Schwarzschild BH value and serves as an excellent measure of BH spin, as it remains largely unaffected by the observer's inclination or the specifics of the BH spacetime geometry. For moderate spin values ($a\lesssim 0.5M$), the rotation parameter values remain consistent across all spacetimes, with constant $\langle \bar{\delta} \rangle_{\psi}$ lines appearing nearly vertical and overlapping across different spacetimes. This consistency makes $\delta$ a sensitive and reliable indicator of BH spin. The red vertical line in each figure identify BHs with $a=0.5M$ and $Q=0.4M$, same as considered in Fig.~\ref{fig:half spin and extra charge}.   }
		\label{fig:KerrNewman Contour Plot}
	\end{figure*}
	
	\begin{figure*}
		\centering
		\includegraphics[width=\linewidth]{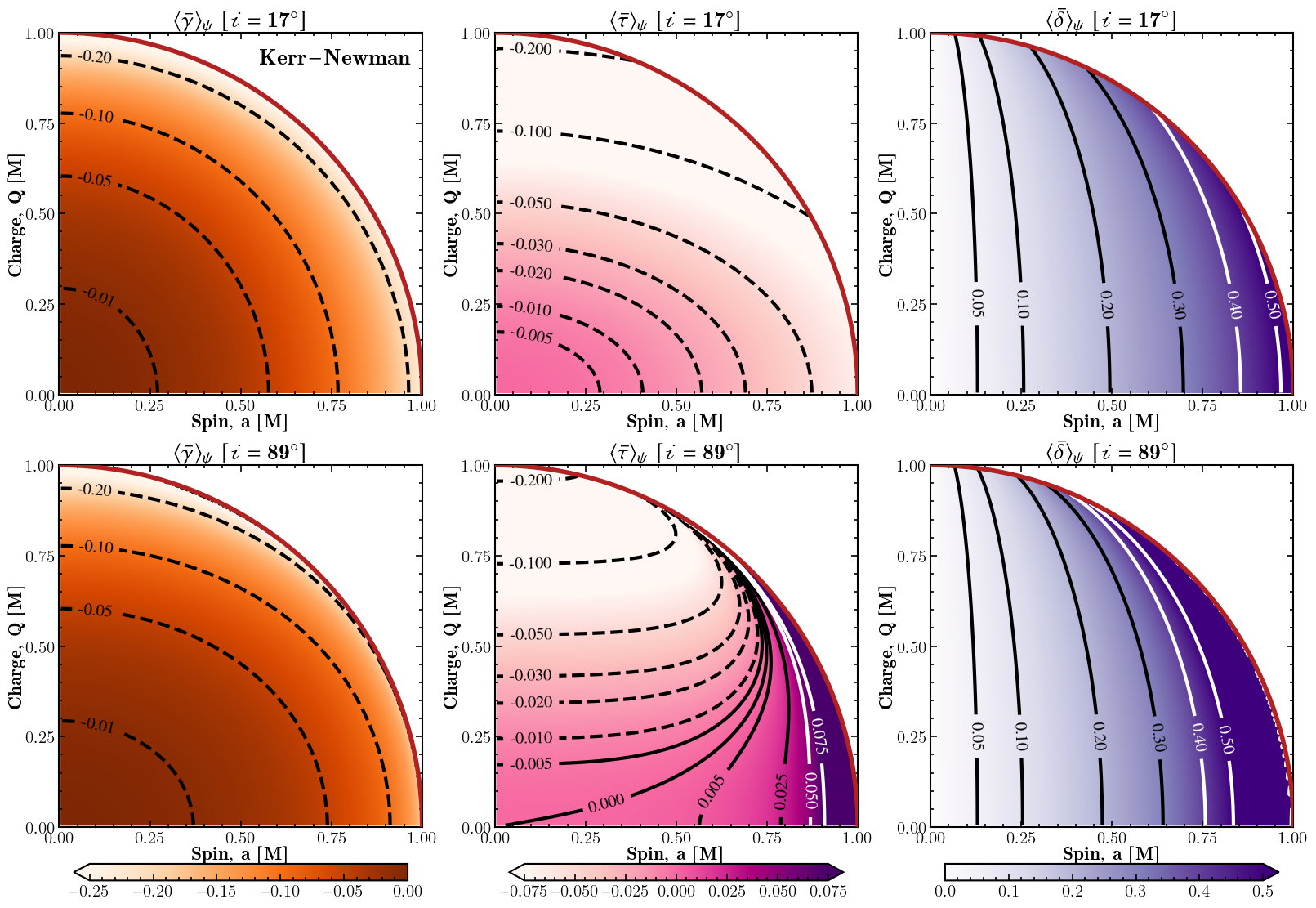}
		\caption{Variation of the fractional deviation in mean critical parameters---demagnification exponent (left column), time delay (center column), and rotation parameter (right column)---for Kerr-Newman BHs with charge and spin, as observed by an observer at a low inclination of $17^\circ$ (top row) and a high inclination of $89^\circ$(bottom row). }
		\label{fig:charge_spin_contour_plot}
	\end{figure*}
	
	The Kerr, Kerr-Newman, and Kerr-Sen BH spacetimes contain curvature singularities in their interiors, whereas the Kerr-Bardeen and Kerr-Hayward BH spacetimes are regular everywhere. Furthermore, barring the Kerr solution, the remaining spacetimes contain self-gravitating matter. 
	The Kerr-Newman spacetime \cite{Newman:1965my} includes an electromagnetic field, the Kerr-Sen spacetime \cite{Sen:1992ua} contains a scalar field, an electromagnetic field, and an axion field, and regular BHs \cite{Bambi:2013ufa} involve exotic matter that violates classical energy conditions deep within their event horizons. In addition to spin ($a$), these BHs also carry a ``generalized charge" ($Q$) due to their nontrivial matter content.
	
	All of these BH models belong to the metric family described by Eq.~\eqref{eq:JP_Metric}, and are thus axisymmetric with integrable geodesic structure. Axisymmetry specifically implies that polar observers (present along the BH spin axis) will find no variations in the critical parameters when moving along the critical curve \cite{Gralla:2020nwp, Salehi:2024cim}. Inclined observers will, however, witness a variation of the critical parameters that depends on the screen angle, $\psi$. The critical parameters are therefore usually functions of $\psi$ for general observer inclinations. 
	
	Future observations, capable of resolving photon subrings, would be able to sample the critical parameters at multiple screen angles. For instance, resolved images of flares caused by orbiting hotspots might enable measurements of all three critical parameters by correlating intensity fluctuations across successive ordered images \cite{Hadar+2021}. In practice, observables would likely be derived from averages of events across multiple occurrences of various phenomena.
	
	Roughly motivated by these expectations, we study the variations in the $\psi-$averaged values of the critical parameters from their Schwarzschild values with varying spacetime geometries and observer inclinations. We discuss the impact of using the $\psi-$median values of the critical parameters as an alternative characteristic measure in Appendix \ref{app:AppB_NonKerr_Param}.
	The specific quantities we study are defined in Eq.~\eqref{eqn:ave_critical_exponents} and Eq.~\eqref{eqn:deviation_critical_exponents}.
	We represent our results through four main studies depicted in Figs.~\ref{fig:half spin and extra charge}, \ref{fig:KerrNewman Contour Plot}, \ref{fig:charge_spin_contour_plot} and \ref{fig:m87_charge_spin_contour_plot} below.
	
	Figure~\ref{fig:half spin and extra charge} shows the variation in the mean values of critical parameters across different spacetimes as a function of observer inclination. Specifically, we present the fractional deviation of these mean values from their corresponding values in a Schwarzschild BH spacetime, emphasizing the influence of nonzero spin and charge with varying inclination. In this figure, the spin and charge of these BHs are fixed at intermediate values of  $a=0.5M$ and $Q=0.4M$, respectively. The variation in all three critical parameters across BHs spacetimes---containing a variety of matter fields---is modest ($\lesssim 5\%$) for all observer inclinations. For the chosen spin and charge, across all studied non-Kerr spacetimes, the maximum deviations observed in the demagnification exponent, the time delay, and the rotation parameter from their corresponding Schwarzschild values are $\lesssim 10\%$, $\lesssim 5\%$, and $\lesssim 25\%$, respectively. This suggests that the demagnification exponent and the rotation parameter are more sensitive probes of the spacetime geometry than the time delay, independently of the observer's inclination. Moreover, for a fixed spin and charge, different spacetimes exhibit distinct deviations in these critical values.
	
	To better understand which aspects of the spacetime each critical parameter is sensitive to, we allow the BH spin to vary in Fig.~\ref{fig:KerrNewman Contour Plot}. We present our analysis for three representative spacetimes: Kerr BHs, Kerr-Newman BHs, and regular Kerr-Bardeen BHs. The BH charge is kept fixed to the previous value of $Q = 0.4M$. Thus, in this figure, the vertical red lines correspond to the BH models shown in Fig.~\ref{fig:half spin and extra charge}. Across the three spacetimes, the mean time delay, $\langle\bar{\tau}\rangle_\psi$, shown in the middle column, features complex behavior with varying BH spin and the observer inclination. Notably, across all three spacetimes and over the entire parameter space of ($a, \mathscr{i}$), we find $\langle\bar{\gamma}\rangle_\psi\leq 0$ and $\langle\bar{\delta}\rangle_\psi\geq 0$, while $\langle\bar{\tau}\rangle_\psi\leq 0$, except for high inclinations or near-extremal BHs. This indicates that a non-Schwarzschild BH can exhibit the same time delay as a Schwarzschild BH, yet differences in the demagnification exponent and rotation parameter can break this degeneracy, allowing these spacetimes to be distinguished. Nevertheless, the time delay, $\langle\bar{\tau}\rangle_\psi$, remains within $\approx 5\%$ of its Schwarzschild value, except in regions of large spins and high inclinations, as seen in the upper-right quadrants of the parameter space. The sharp increase in time delay at large spins and high inclinations---particularly for near-extremal BHs---arises from contributions from photon trajectories approaching the BH event horizon, where gravitational time dilation becomes significant. For an extremal ($a=M$) Kerr BH in particular, Eq.~\eqref{eq:thetaTurning} can be used to show that photons with trajectories close to the horizon also remain close to the equatorial plane, and are, thus, only seen by observers located close to the equatorial plane. In contrast, for typical (nonextremal) BHs, the time delay is relatively insensitive to the observer’s inclination. Moreover, for small inclinations, the time delay varies by approximately 8\% with spin---similar to the variation in shadow size (see Eq.~(45) and Fig.~1 in Ref.~\cite{Salehi:2024cim}).

	In contrast to the time delay, the demagnification exponent (left column) and the rotation parameter (right column) have simpler behavior. The demagnification exponent is smaller and lies within about $20\%$ of, its Schwarzschild value in all of these BH spacetimes. A smaller demagnification exponent indicates that photon subrings in these non-Kerr spacetimes will have larger widths and fluxes as compared to those in Schwarzschild BH spacetime.
	
	The rotation parameter, on the other hand, is always found to be larger than its Schwarzschild value and is effectively insensitive to the observer's inclination for moderate BH spin. This behavior was expected from the spin-induced frame-dragging contribution in the rotation parameter for spinning BHs. Furthermore, comparing isocontours of the same rotation parameter value across the different BH spacetimes shows that they appear at the same BH spin values. These findings collectively suggest that the rotation parameter may be an excellent proxy for BH spin, independently of the observer inclination and the specifics of the spacetime geometry.
	
	\begin{figure}
		\centering
		\includegraphics[width=\linewidth]{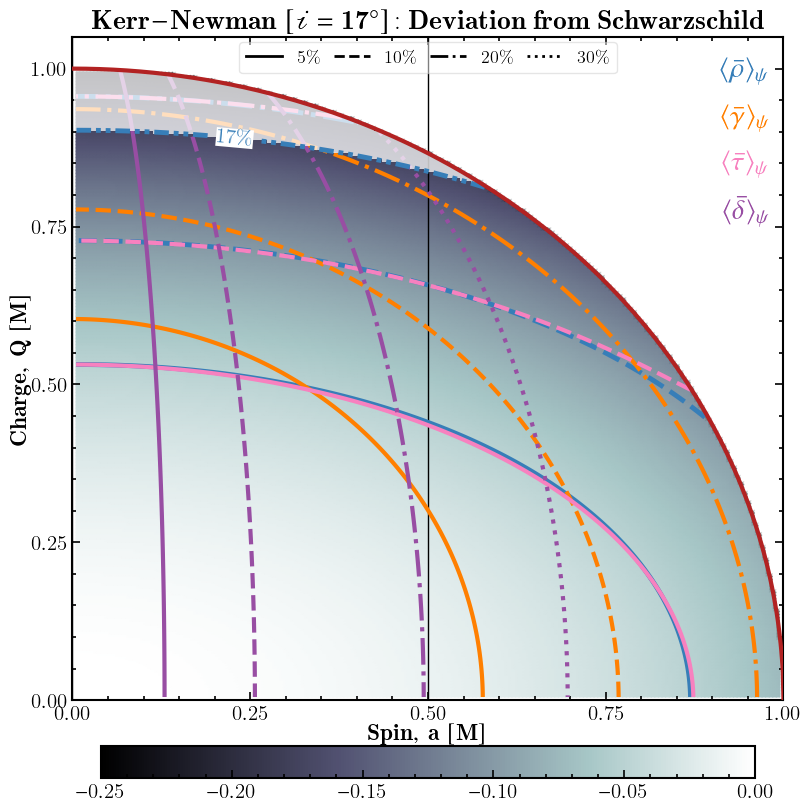}
		\caption{Combining measurements of various critical parameters could pin down physical properties of astrophysical BHs. Shown here are the deviations in the shadow size (blue lines and background color), the demagnification exponent (orange), the time delay (pink), and the rotation parameter (purple) of a Kerr-Newman BH from their Schwarzschild values. The isocontours display 5\% (solid), 10\% (dashed), 20\% (dot-dashed) and 30\% (dotted) deviations. The white-shaded region is excluded by the 2017 EHT shadow size measurement of M87$^*$. The isocontours of shadow size and time delay are virtually indistinguishable.}
		\label{fig:m87_charge_spin_contour_plot}
	\end{figure}
	
	To investigate whether the rotation parameter is truly a measure of the BH spin alone, we finally allow the BH charge of Kerr-Newman BH to vary in Fig.~\ref{fig:charge_spin_contour_plot}. As a reminder, the zero charge limit of the Kerr-Newman spacetime yields the Kerr spacetime. We show the fractional deviation in mean rotation parameter in the right column as seen by an observer close to the BH spin axis at an inclination of $17^\circ$ (top row) or located close to the equatorial plane ($\mathscr{i} = 89^\circ$; bottom row).  The former inclination was chosen to be consistent with the large scale jet of M87$^*$ \cite{Walker+2018}, and the latter was chosen to be exemplary of cases with extremely high inclinations. These panels provide further evidence that the rotation parameter is typically an excellent measure of the BH spin, since the isocontours lines appear to be nearly parallel to the charge axis. Close to the extremal limit (circular red line), however, the rotation parameter does become sensitive to the BH charge. 
	
	We interpret the sensitivity of the rotation parameter to spin alone as a direct manifestation of its role in measuring the magnitude of frame-dragging, or gravitomagnetism, experienced by nearly bound photon orbits. Notably, the first term in $\delta_{\rm p}$ in Eq.~(\ref{Eq:Critical-parameter}) varies linearly with spin. Furthermore, if such measurement yields a nonzero value in particular, $\langle\bar{\delta}\rangle_\psi\neq 0$ it would correspond to a measurement of frame-dragging close to a BH horizon, directly analogous to a measurement of the gravitomagnetic precession of the Gravity Probe B gyroscopes in the Earth's gravitational field or that measured in the double pulsar system \cite{Breton+2008}.
	
	The demagnification exponent (left column) shows almost equal sensitivity to both BH spin and charge, for this BH spacetime. 
	
	Finally, the time delay exhibits smooth variations with spin and charge at small inclinations (top middle panel), showing a relatively higher sensitivity to the BH charge while remaining within approximately $\lesssim 20\%$ of the Schwarzschild value.  At high inclinations (bottom middle panel),  the influence of nearly equatorial photon orbits becomes evident, particularly in near-extremal BHs, where these orbits approach the event horizon. These photon orbits, tracing the near-horizon spacetimes, experience substantial time dilation, which delays the formation of certain segments of the photon ring, ultimately influencing its overall mean value.
	
	In Fig.~\ref{fig:m87_charge_spin_contour_plot}, we consolidate our findings into a predictive framework for future high-resolution imaging of M87$^*$. This figure displays isocontours of the fractional deviation in the mean values of critical parameters for Kerr-Newman BHs as functions of spin and charge, for an observer at inclination of $17^\circ$. We additionally highlight the fractional deviation in the mean shadow size, $\langle \bar{\rho} \rangle_\psi$, of Kerr-Newman BHs from its Schwarzschild value ($\sqrt{27}M$) using blue isocontours. The isocontours represent absolute deviations of $5\%$ (solid), $10\%$ (dashed), $20\%$ (dot-dashed) and $25\%$ (dotted) from the Schwarzschild values. The white-shaded region delineates the constraints on the Kerr-Newman BH parameter space based on the 2017 EHT shadow size measurement of M87$^*$, as detailed in Refs.~\cite{Psaltis+2020, Kocherlakota+2021}. These indicative constraints were obtained by determining the deviation in the mass inferred from the emission ring in the image of M87$^*$ by the EHT \cite{EHTC+2019f}, compared to that from stellar dynamics measurements \cite{Gebhardt+2011}.
	
	This figure clearly demonstrates that the isocontours of $\langle \bar{\rho} \rangle_\psi$ and $\langle \bar{\tau} \rangle_\psi$ are nearly indistinguishable, effectively overlapping. Consequently, the mean time delay serves as an independent estimator of shadow size, particularly at low observer inclinations (see Eq.~(43) and Fig.~1 of Ref.~\cite{Salehi:2024cim} for details). In observations of M87*, where the jet orientation suggests a modest inclination angle relative to the observer, this result is particularly advantageous: a high temporal resolution measurement of time delays from transient events, such as flaring, offers an independent and complementary estimate of the shadow size. Furthermore, this figure illustrates the power of joint measurements of the photon ring critical parameters to pin down physical properties of astrophysical BHs, \textit{viz.}, their spin and charge. Joint measurements of either shadow size or time delay, along with the demagnification exponent and rotation parameter, give precise estimations of both BH spin and generalized charge.
	
\begin{figure}
		\centering
		\includegraphics[width=\linewidth]{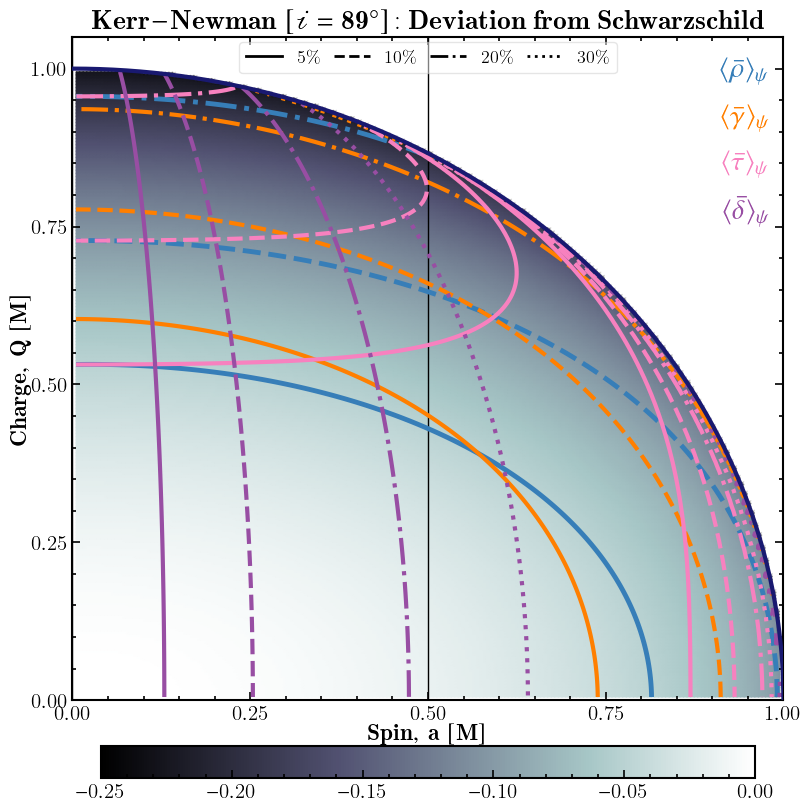}
		\caption{Similar plot to Fig.~\ref{fig:m87_charge_spin_contour_plot} but for an observer present close to the equatorial plane. Notably, the time delay and the shadow size display distinct behaviors for Kerr-Newman BHs of different spin and charge for general observer inclinations. Nevertheless, the joint measurement of the shadow size and either of these critical parameters can precisely determine the BH spin and charge.}
		\label{fig:Equatorial Delay Time and Shadow Size}
	\end{figure}

	Finally, to assess whether the time delay consistently provides a reliable measure of the shadow size, in Fig.~\ref{fig:Equatorial Delay Time and Shadow Size}, we replicate Fig.~\ref{fig:m87_charge_spin_contour_plot} but change the observer inclination to $\mathscr{i} = 89^\circ$. It is immediately evident that both quantities, $\langle \bar{\rho} \rangle_\psi$ and $\langle \bar{\tau} \rangle_\psi$, exhibit distinct behaviors with varying BH spin and charge, and that they conspire to produce identical variations only for observers present close to the BH spin axis. Note that two isocontours for $5\%$ deviation in time delay (solid curves) correspond to $\langle \bar{\tau} \rangle_\psi =\pm 0.05$.
	
	\section{Summary And Conclusions}
	\label{Sec-4}
	
	Horizon-scale images of BHs have opened a new frontier for probing strong-field gravity and revealing key properties of astrophysical BHs, such as their spin. A particularly exciting prospect for future BH imaging, with better angular resolution and greater flux sensitivity, is the direct detection of the BH photon ring. Due to strong gravitational lensing, a single point source can produce multiple images on the observer's screen. The primary image is the first to arrive on the screen, while ``higher-order'' images---collectively known as photon ring for an extended optically thin source of emission---emerge later. These increasingly higher-order images are formed by photons that execute multiple (half-)orbits around the BH, carrying unique signatures of the near-horizon spacetime geometry.
	
	The novelty of this work lies in extending the analyses of Refs.~\cite{Gralla:2019drh} and \cite{Salehi+2023}. While Ref.~\cite{Gralla:2019drh} derives critical parameters exclusively for Kerr BHs and Ref.~\cite{Salehi+2023} examines only the shadow size and Lyapunov exponent for a polar observer, this study explores how strong-field deviations from the Kerr spacetime affect all critical parameters across varying inclinations.
	
	Our investigation focuses on three critical parameters that characterize the universal relations between higher-order images: the demagnification exponent ($\gamma$) governs the demagnification of successive images, the time delay ($\tau$) determines the elapsed time between their appearance, and the rotation parameter ($\delta$) relates their angular (azimuthal) positions on the image plane. We present a comprehensive analysis of how these parameters vary across a diverse family of non-Kerr spacetimes---including the Kerr-Newman, Kerr-Sen, Kerr-Bardeen, and Kerr-Hayward metrics---as functions of BH spin ($a$), generalized charge ($Q$), and the observer's inclination ($\mathscr{i}$). Specifically, we quantify the fractional deviations of the azimuthally averaged values of these critical parameters ($\langle \bar{\gamma} \rangle_\psi$, $\langle \bar{\tau} \rangle_\psi$, $\langle \bar{\delta} \rangle_\psi$) from those of a Schwarzschild BH ($\pi, \sqrt{27}M \pi, \pi$). Our key findings are summarized below.
	
	The critical parameters respond distinctively to variations in $a$, $Q$, and $\mathscr{i}$, establishing them as complementary probes of spacetime geometry. Specifically, the demagnification exponent is always smaller in non-Schwarzschild spacetimes (including Kerr) as compared to its Schwarzschild BH value. This indicates that, for primary images of the same size, secondary images are always larger in non-Schwarzschild spacetimes. The time delay for non-Schwarzschild BHs is smaller than its Schwarzschild BH value, except for near-extremal BHs viewed from high inclinations for which $\langle \bar{\tau} \rangle_\psi\geq 0$. For polar and near-polar observers, the time delay serves as an excellent proxy for the BH shadow size. Finally, the rotation parameter is always larger in non-Schwarzschild spacetimes (including Kerr) and serves as an excellent measure of BH spin. It is essentially insensitive to the BH charge, the specifics of the spacetime geometry (e.g., a Kerr-Newman vs. a Kerr-Bardeen of the same spin and charge), and the observer's inclination. 
	
	Thus, $\langle \bar{\gamma} \rangle_{\psi}\leq 0$, $\langle \bar{\delta} \rangle_{\psi}\geq 0$, and $\langle \bar{\tau} \rangle_\psi\leq 0$ imply that photon subrings in these non-Kerr spacetimes are wider, brighter, more rotated, and advanced in time compared to those in the Schwarzschild spacetime.
	
	Quantitatively, for Kerr BHs, the fractional deviations across the full parameter space of spin and inclination are: $\sim 20\% $ in $\langle \bar{\gamma} \rangle_{\psi}$,  $\sim 10\%$ in $\langle \bar{\tau} \rangle_{\psi}$,  and $\sim 60\%$ in $\langle \bar{\delta} \rangle_{\psi}$---significantly larger than the $8\% $ variation in the shadow radius from their corresponding values for Schwarzschild BH.
	
	While both $\gamma$ and $\delta$ are nearly independent of inclination, $\tau$ is somewhat sensitive to $\mathscr{i}$. In particular, the time delay rises sharply for near-extremal BHs viewed at high-inclinations, due to contributions from nearly-bound orbits that approach the close vicinity of the event horizon. The portion of the photon ring constructed by prograde photons tracing the near-horizon region and experiencing significant time dilation makes the dominant contribution when calculating the average in Eq.~(\ref{eqn:ave_critical_exponents}).
	
	A key finding of our analysis demonstrates that the time delay between successive photon rings serves as an independent measure of shadow size for polar and nearly-polar observer, even for spinning BHs. Our results establish that photon ring critical parameters can effectively constrain both BH spin and non-Kerr deviations. Using the Kerr-Newman metric as an example, we demonstrate that combining shadow size measurements with either the  demagnification exponent or rotation parameter successfully breaks the measurement degeneracy between spin and non-Kerr parameters, i.e., generalized charge. This remains true for a large family of non-Kerr BHs. 
	
	Building on this, we note that each critical parameter offers an independent measurement of a distinct spacetime property near the BH photon shell. The demagnification exponent is sensitive to spacetime curvature, the time delay reflects the gravitational potential governing time dilation, and the rotation parameter provides insight into spacetime frame-dragging. Therefore, joint measurements of these critical parameters can help deduce key BH properties.
	
	Encouragingly, with the growing capabilities of interferometric arrays, such as ngEHT \cite{Johnson+2023} and BHEX \cite{Johnson:2024ttr}, the precision required to detect $n=1$ photon ring is now within reach, bringing us closer to addressing fundamental questions about the nature of BHs and their potential deviations from the Kerr spacetimes.
	
	Nonetheless, while critical parameters provide a powerful framework for characterizing photon ring and BH features, potential degeneracies between different spacetime solutions may limit their ability to uniquely identify the underlying spacetime geometry and deviations (e.g., charge) from the Kerr geometry. 
	
	Motivated by these findings, this work opens new research directions: (i) assessing the accuracy of critical parameters in defining the scaling relations between $n=0$ and $n=1$ rings, (ii) how precisely the calibrated critical parameters, that define the scaling relations between $n=0$ and $n=1$ rings, determine the BH parameters. Our future research is exploring these questions.

	
	\begin{acknowledgments}
		R.K.W.'s research is supported by the Fulbright-Nehru Postdoctoral Research Fellowship (Award No. 2847/FNPDR/2022) from the United States-India Educational Foundation. 
		P.K. acknowledges support from grants from the Gordon and Betty Moore Foundation (No.~GBMF-8273) and the John Templeton Foundation (No.~62286) to the Black Hole Initiative at Harvard University. 
		D.O.C acknowledges financial support from the Brinson Foundation, the Gordon and Betty Moore Foundation (No.~GBMF-10423), and the National Science Foundation (No.~AST-2307887, No.~AST-1935980, and No.~AST-2034306). 
		K.S. acknowledges support from the Perimeter Institute for Theoretical Physics. Funding for research at the institute is provided by the Department of Innovation, Science and Economic Development Canada, and the Ministry of Economic Development, Job Creation and Trade of Ontario, both of which are branches of the Government of Canada. 
		The opinions expressed in this work are those of the authors and do not necessarily reflect the views of these foundations.
	\end{acknowledgments}

	\begin{appendix}
		
		\section{Representative Measures of The critical Parameters}
		\label{app:AppA_Measures}

		\begin{figure*}
			\centering
			\includegraphics[width=\linewidth]{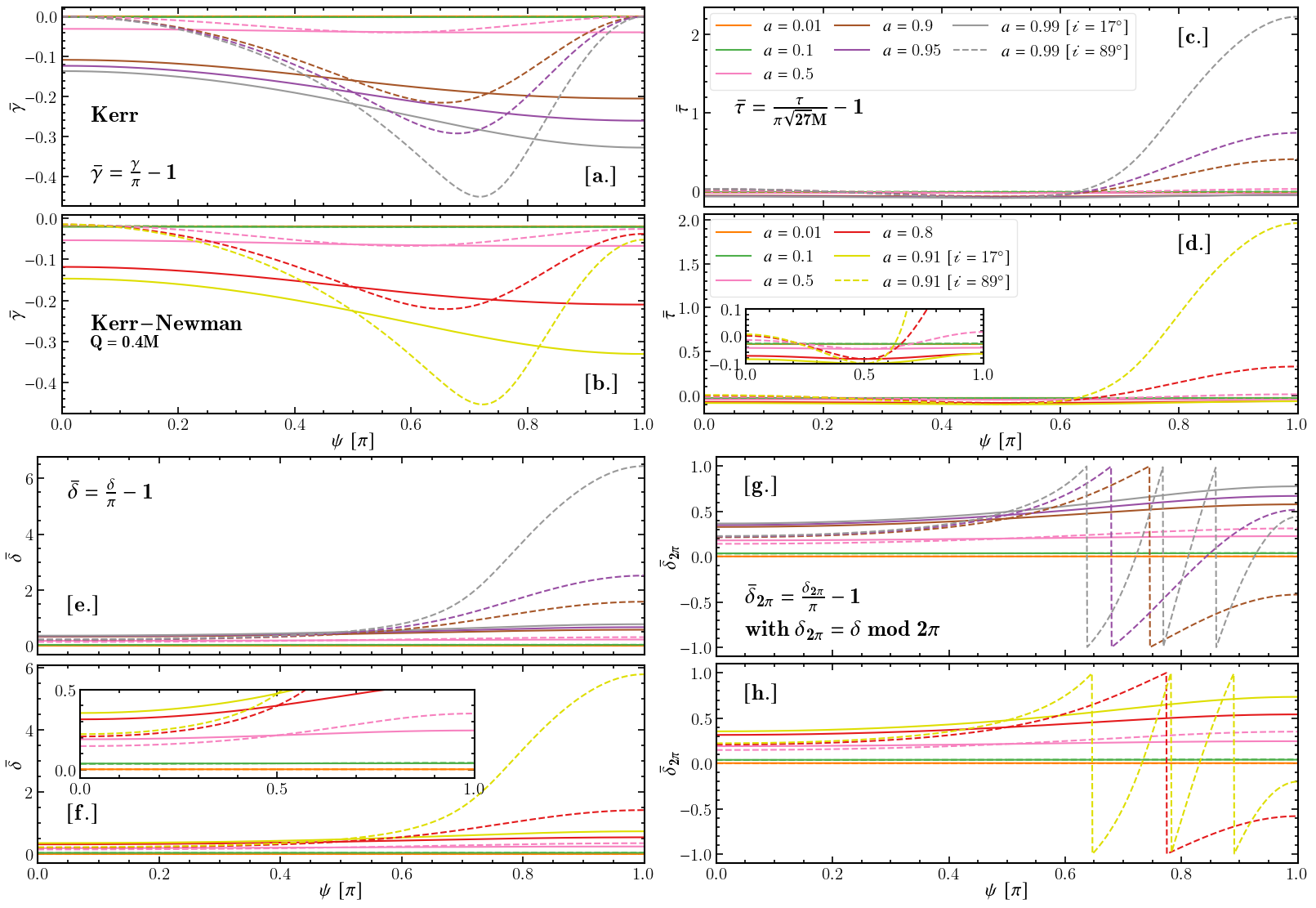}
			\caption{Variation of fractional deviation in critical parameters for the Kerr BH [panels (a), (c), (e), (g)] and Kerr-Newman BH with $Q=0.4$ [panels (b), (d), (f), (h)]. All deviations are shown with respect to their Schwarzschild values over varying image plane polar angle $\psi$; the demagnification exponent, $\bar\gamma$, [panels (a) and (b)], the time delay between successive images, $\bar\tau$, [(c) and (d)], the rotation parameter, $\bar\delta$, [(e) and (f)], and the phase wrapped rotation parameter, $\bar\delta_{2\pi}$, [(g) and (h)]. We show deviation curves for varying spin values varying from $0.01\%$ to $99\%$ of extremality for spacetimes in varying colors, and for observer inclination angles of an M87$^*$-like system of $17^\circ$ (solid), and for an extreme inclination case of $89^\circ$ (dashed). The insets of (d) and (f) displayed a zoomed in version of their encompassing panels.  }    \label{fig:Kerr_KerrNewman_critical_exponent_variation}
		\end{figure*}
		
		\begin{figure*}
			\centering
			\includegraphics[width=\linewidth]{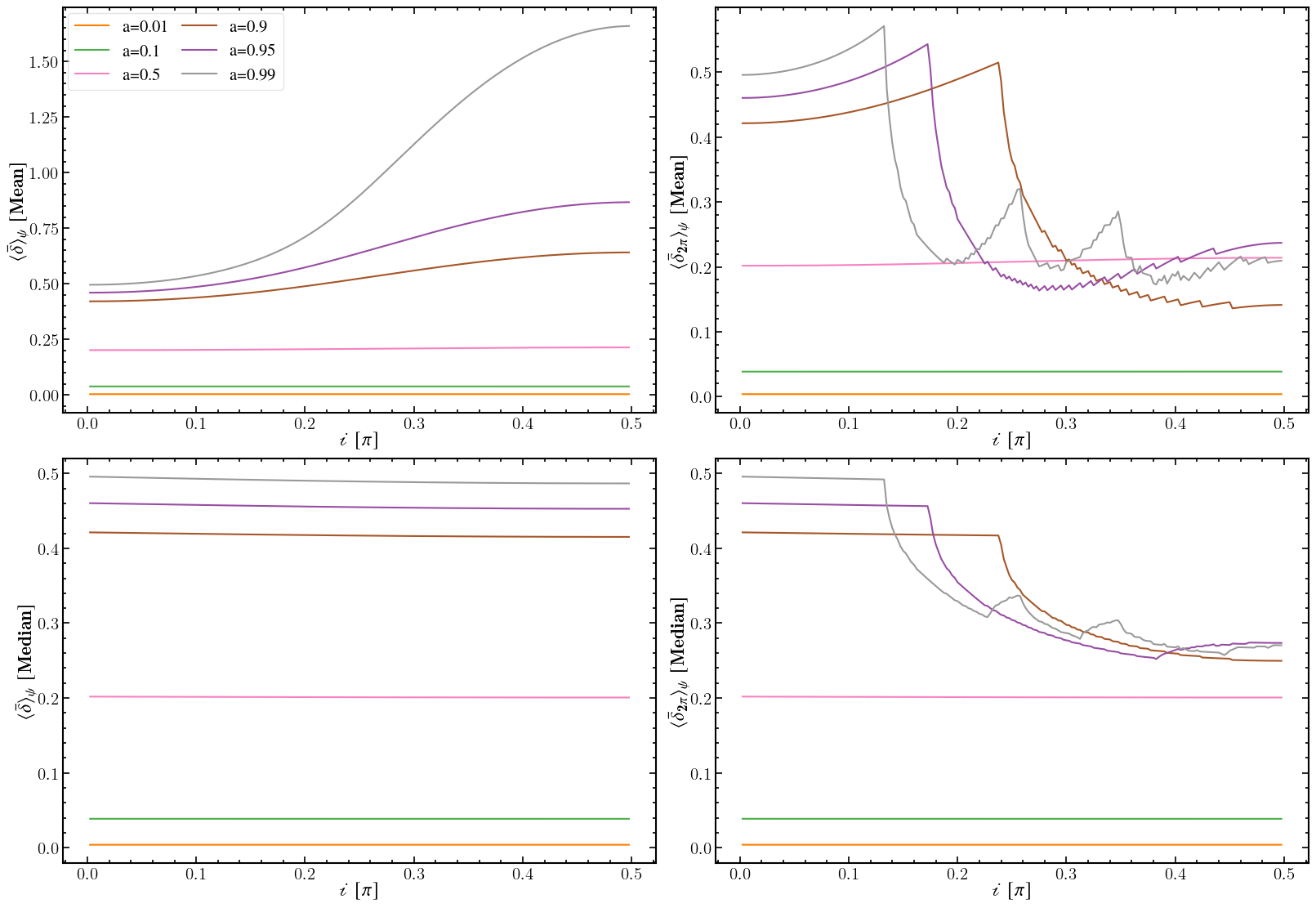}
			\caption{Figure illustrating different statistical measures of the critical rotation parameter along the photon ring.
				We show the un-phase wrapped statistics in the left column, and the phased wrapped statistics in the right column, where we have applied a $2\pi$ modulo operation before constructing the statistics. We specifically show the screen averaged mean (top row) and median (bottom row) values. Whether the $2\pi$ modulo operation must be taken before the statistic is dependent on the nature of the underlying observation (See the discussion at the end of Sec.~\ref{Sec-2}). We adopt the mean value as our primary measure, although they vary more over observer inclination than their corresponding median values, especially at high values of inclinations. Unlike the median, which remains largely insensitive to inclination (see bottom-left sub-figure), the mean captures the inclination's imprint and varies smoothly with it. Additionally, averages having modulo operation after statistical averaging, exhibit smooth variations with inclination angle}
			\label{fig:critical_exponent_mean_and_median}
		\end{figure*}
		
		For spinning BHs, the observer's inclination angle, $\mathscr{i}$, determines the part of photon shell, $r_{\mathrm{p}}^- \leq r \leq r_{\mathrm{p}}^+$, satisfying $\Theta(\mathscr{i}) \geq 0$, that contributes to the image formation on the observer's screen. This leads to a rich phenomenology where polar observers ($\mathscr{i}=0,\pi$) see perfectly circular photon subrings formed exclusively by photons on the polar SNG at radius $r_{\rm p}^0$, while equatorial observers access the complete photon shell, resulting in maximal image distortion, with smooth variation at intermediate inclinations. As the image plane polar coordinate $\psi$ varies along a photon subring or the critical curve, we trace SNGs of different BL radii within the visible portion of the photon shell, transitioning from retrograde orbits at $\psi = 0$ to prograde orbits at $\psi = \pi$.
		
		These critical parameters (\ref{Eq:Critical-parameter}), functions of SNG radius $r_{\rm p}$, can be expressed in terms of $\psi$ using Eq.~(\ref{Eq:rp-psi}). For polar inclination angles, these parameters remain constant along the photon ring. However, for non-polar inclinations ($\mathscr{i}\neq 0,\pi$), they vary with $\psi$, causing distinct portions of the photon ring to be demagnified, rotated, and delayed by varying amounts (see Fig.~\ref{fig:Kerr_KerrNewman_critical_exponent_variation}).
		
		Figure \ref{fig:Kerr_KerrNewman_critical_exponent_variation} shows the fractional deviation of critical parameters from their Schwarzschild values as a function of image plane polar angle $\psi$ for both Kerr and Kerr-Newman BHs. We analyze these deviations at two distinct inclination angles ($17^\circ$ and $89^\circ$) across various spin parameters. The results demonstrate that critical parameters in spinning BHs exhibit $\psi$-dependent variations, with this effect becoming particularly pronounced at higher inclinations.
		
		While $\bar{\gamma}<0$ for all values of $a$ and $\mathscr{i}$, at high inclinations, $\bar{\gamma}$ exhibits nonmonotonic variation between $\psi=0$ and $\psi=\pi$, reaching a minimum at an intermediate value. This minimum in $\bar{\gamma}$ corresponds to the screen angle where the photon ring appears widest and brightest. Both $\bar{\delta}\geq 0$ and $\bar{\tau}\geq 0$ further increase with the screen angle $\psi$. This monotonically increasing time delay with $\psi$ is an indicator that the distinct part of a given order photon ring takes distinct time to appear. This temporal variation arises because prograde photons $r_{\rm p}\leq r_{\rm p}^0$ forming a part of photon ring takes longer to reach the observer' screen due to time dilation, compared to their retrograde counterparts. Consequently, for a counterclockwise spinning BH, the left part of the photon ring ($\pi/2\leq \psi\leq 3\pi/2$) on the image plane appears more wider and brighter, more rotated, and delayed than its right part.
		
		For BHs with large spin parameter and high inclination angle,  photon ring segments formed by prograde photons tracing the near-horizon region and experiencing strong frame-dragging can undergo rotations exceeding $2\pi$ radians. This is illustrated in the lower panel of Fig.~\ref{fig:Kerr_KerrNewman_critical_exponent_variation}, where the rotation parameter $\delta\geq 2\pi$. When imaging static sources, the wrapped measured rotation parameter $\delta_{2\pi}$ displays a zigzag pattern.
		
		Time-resolved observations of dynamic emission sources around BHs could potentially reveal both temporal variations in photon ring formation and segments rotated beyond $2\pi$. These phenomena present compelling targets for future time-domain observations of BH dynamics.
		
		For quantitative analysis, statistical measures of the critical parameter along the photon ring are required. We made four distinct measurements of the rotation parameter: (i) median[mod($\bar{\delta}, 2\pi$)], (ii) mean[mod($\bar{\delta}, 2\pi$)], (iii) mod[median($\bar{\delta}$), $2\pi$], and (iv) mod[mean($\bar{\delta}$), $2\pi$]. Figure \ref{fig:critical_exponent_mean_and_median} presents these measurements for a Kerr BH as a function of spin with varying inclination. Measurements (iii) and (iv), which apply the modulo operation after statistical averaging, exhibit smooth variations with inclination angle, whereas measurements (i) and (ii) show multiple discontinuities, suggesting measurement artifacts. Based on these results, we adopt mod[mean($\bar{\delta}$), $2\pi$] for our rotation parameter calculations.

		\section{Variation with BH Charge}
		\label{app:AppB_NonKerr_Param}
		\begin{figure*}
			\centering
			\includegraphics[width=\linewidth]{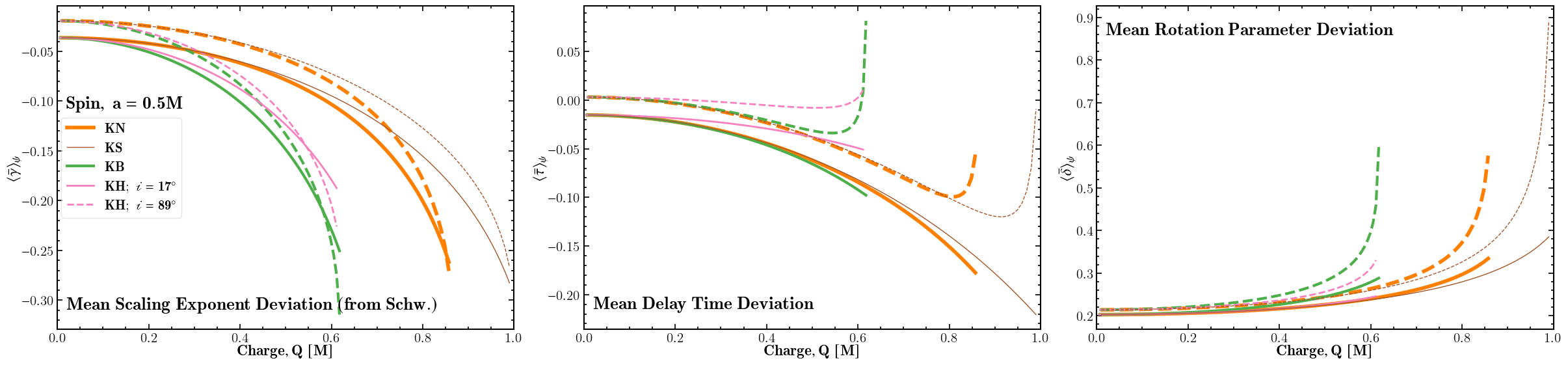}
			\caption{Variation of fractional deviation of critical parameters with charge for non-Kerr BHs. The fractional deviation of the mean values of the demagnification exponent and time delay from their Schwarzschild counterparts decreases and increases monotonically, respectively, with increasing charge. For non-extremal BHs, the time delay decreases with charge. However, for large inclinations and near-extremal BHs, it sharply increases due to photons traversing paths very close to the extremal horizon.}\label{fig:charge-critical_exponent}
		\end{figure*}
		
		Non-Kerr BHs considered in this paper are characterized by a charge parameter $Q$ in addition to mass $M$ and spin $a$. The influence of charge on BH shadow morphology has been extensively studied in both spherically symmetric and rotating spacetimes, with corresponding observational constraints \cite{Kocherlakota+2021}.
		
		Figure~\ref{fig:charge-critical_exponent} illustrates the charge dependence of photon ring critical parameters across different BH spacetimes at observer's inclination of $17^\circ$ and $89^\circ$. For a fixed inclination and spin, increasing the BH charge produces several distinct effects: the demagnification exponent decreases, the rotation parameter increases, and the time delay monotonically decreases at small inclination angles. However, for large inclinations, the time delay initially decreases with increasing charge but then sharply rises for near-extremal value of charge, eventually exceeding the Schwarzschild BH value. This behavior underscores the significant time dilation experienced by prograde orbits near the equatorial plane and close to the BH horizon. Consequently, for observers at high inclinations, a measured time delay value may correspond to two BH spacetimes within the same family that have identical spins but distinct charges. Notably, for all BH solutions examined, we find $\langle\bar{\gamma}\rangle < 0$ and $\langle\bar{\delta}\rangle > 0$ across all charge values, with maximal effects manifesting in near-extremal configurations.
		
		Among all observables studied, the demagnification exponent governing ring demagnification demonstrates maximum charge sensitivity, whereas time delay exhibits minimal charge dependence.
	\end{appendix}

	\bibliography{PRD_Refs.bib}

\providecommand{\noopsort}[1]{}\providecommand{\singleletter}[1]{#1}%
\begin{thebibliography}{69}%
\makeatletter
\providecommand \@ifxundefined [1]{%
 \@ifx{#1\undefined}
}%
\providecommand \@ifnum [1]{%
 \ifnum #1\expandafter \@firstoftwo
 \else \expandafter \@secondoftwo
 \fi
}%
\providecommand \@ifx [1]{%
 \ifx #1\expandafter \@firstoftwo
 \else \expandafter \@secondoftwo
 \fi
}%
\providecommand \natexlab [1]{#1}%
\providecommand \enquote  [1]{``#1''}%
\providecommand \bibnamefont  [1]{#1}%
\providecommand \bibfnamefont [1]{#1}%
\providecommand \citenamefont [1]{#1}%
\providecommand \href@noop [0]{\@secondoftwo}%
\providecommand \href [0]{\begingroup \@sanitize@url \@href}%
\providecommand \@href[1]{\@@startlink{#1}\@@href}%
\providecommand \@@href[1]{\endgroup#1\@@endlink}%
\providecommand \@sanitize@url [0]{\catcode `\\12\catcode `\$12\catcode
  `\&12\catcode `\#12\catcode `\^12\catcode `\_12\catcode `\%12\relax}%
\providecommand \@@startlink[1]{}%
\providecommand \@@endlink[0]{}%
\providecommand \url  [0]{\begingroup\@sanitize@url \@url }%
\providecommand \@url [1]{\endgroup\@href {#1}{\urlprefix }}%
\providecommand \urlprefix  [0]{URL }%
\providecommand \Eprint [0]{\href }%
\providecommand \doibase [0]{https://doi.org/}%
\providecommand \selectlanguage [0]{\@gobble}%
\providecommand \bibinfo  [0]{\@secondoftwo}%
\providecommand \bibfield  [0]{\@secondoftwo}%
\providecommand \translation [1]{[#1]}%
\providecommand \BibitemOpen [0]{}%
\providecommand \bibitemStop [0]{}%
\providecommand \bibitemNoStop [0]{.\EOS\space}%
\providecommand \EOS [0]{\spacefactor3000\relax}%
\providecommand \BibitemShut  [1]{\csname bibitem#1\endcsname}%
\let\auto@bib@innerbib\@empty
\bibitem [{\citenamefont {Akiyama}\ \emph
  {et~al.}(2019{\natexlab{a}})\citenamefont {Akiyama} \emph
  {et~al.}}]{EHTC+2019a}%
  \BibitemOpen
  \bibfield  {author} {\bibinfo {author} {\bibfnamefont {K.}~\bibnamefont
  {Akiyama}} \emph {et~al.} (\bibinfo {collaboration} {Event Horizon
  Telescope}),\ }\bibfield  {title} {\bibinfo {title} {{First M87 Event Horizon
  Telescope Results. I. The Shadow of the Supermassive Black Hole}},\ }\href
  {https://doi.org/10.3847/2041-8213/ab0ec7} {\bibfield  {journal} {\bibinfo
  {journal} {Astrophys. J. Lett.}\ }\textbf {\bibinfo {volume} {875}},\
  \bibinfo {pages} {L1} (\bibinfo {year} {2019}{\natexlab{a}})},\ \Eprint
  {https://arxiv.org/abs/1906.11238} {arXiv:1906.11238 [astro-ph.GA]}
  \BibitemShut {NoStop}%
\bibitem [{\citenamefont {Akiyama}\ \emph
  {et~al.}(2022{\natexlab{a}})\citenamefont {Akiyama} \emph
  {et~al.}}]{EHTC+2022a}%
  \BibitemOpen
  \bibfield  {author} {\bibinfo {author} {\bibfnamefont {K.}~\bibnamefont
  {Akiyama}} \emph {et~al.} (\bibinfo {collaboration} {Event Horizon
  Telescope}),\ }\bibfield  {title} {\bibinfo {title} {{First Sagittarius A*
  Event Horizon Telescope Results. I. The Shadow of the Supermassive Black Hole
  in the Center of the Milky Way}},\ }\href
  {https://doi.org/10.3847/2041-8213/ac6674} {\bibfield  {journal} {\bibinfo
  {journal} {Astrophys. J. Lett.}\ }\textbf {\bibinfo {volume} {930}},\
  \bibinfo {pages} {L12} (\bibinfo {year} {2022}{\natexlab{a}})},\ \Eprint
  {https://arxiv.org/abs/2311.08680} {arXiv:2311.08680 [astro-ph.HE]}
  \BibitemShut {NoStop}%
\bibitem [{\citenamefont {Akiyama}\ \emph
  {et~al.}(2024{\natexlab{a}})\citenamefont {Akiyama} \emph
  {et~al.}}]{EventHorizonTelescope:2024dhe}%
  \BibitemOpen
  \bibfield  {author} {\bibinfo {author} {\bibfnamefont {K.}~\bibnamefont
  {Akiyama}} \emph {et~al.} (\bibinfo {collaboration} {Event Horizon
  Telescope}),\ }\bibfield  {title} {\bibinfo {title} {{The persistent shadow
  of the supermassive black hole of M87. I. Observations, calibration, imaging,
  and analysis}},\ }\href {https://doi.org/10.1051/0004-6361/202347932}
  {\bibfield  {journal} {\bibinfo  {journal} {Astron. Astrophys.}\ }\textbf
  {\bibinfo {volume} {681}},\ \bibinfo {pages} {A79} (\bibinfo {year}
  {2024}{\natexlab{a}})}\BibitemShut {NoStop}%
\bibitem [{\citenamefont {{Narayan}}\ \emph {et~al.}(2019)\citenamefont
  {{Narayan}}, \citenamefont {{Johnson}},\ and\ \citenamefont
  {{Gammie}}}]{Narayan+2019}%
  \BibitemOpen
  \bibfield  {author} {\bibinfo {author} {\bibfnamefont {R.}~\bibnamefont
  {{Narayan}}}, \bibinfo {author} {\bibfnamefont {M.~D.}\ \bibnamefont
  {{Johnson}}},\ and\ \bibinfo {author} {\bibfnamefont {C.~F.}\ \bibnamefont
  {{Gammie}}},\ }\bibfield  {title} {\bibinfo {title} {{The Shadow of a
  Spherically Accreting Black Hole}},\ }\href
  {https://doi.org/10.3847/2041-8213/ab518c} {\bibfield  {journal} {\bibinfo
  {journal} {apjl}\ }\textbf {\bibinfo {volume} {885}},\ \bibinfo {eid} {L33}
  (\bibinfo {year} {2019})},\ \Eprint {https://arxiv.org/abs/1910.02957}
  {arXiv:1910.02957 [astro-ph.HE]} \BibitemShut {NoStop}%
\bibitem [{\citenamefont {{Bronzwaer}}\ \emph {et~al.}(2021)\citenamefont
  {{Bronzwaer}} \emph {et~al.}}]{Bronzwaer+2021}%
  \BibitemOpen
  \bibfield  {author} {\bibinfo {author} {\bibfnamefont {T.}~\bibnamefont
  {{Bronzwaer}}} \emph {et~al.},\ }\bibfield  {title} {\bibinfo {title}
  {{Visibility of black hole shadows in low-luminosity AGN}},\ }\href
  {https://doi.org/10.1093/mnras/staa3430} {\bibfield  {journal} {\bibinfo
  {journal} {mnras}\ }\textbf {\bibinfo {volume} {501}},\ \bibinfo {pages}
  {4722} (\bibinfo {year} {2021})},\ \Eprint {https://arxiv.org/abs/2011.00069}
  {arXiv:2011.00069 [astro-ph.HE]} \BibitemShut {NoStop}%
\bibitem [{\citenamefont {{Kocherlakota}}\ and\ \citenamefont
  {{Rezzolla}}(2022)}]{Kocherlakota+2022}%
  \BibitemOpen
  \bibfield  {author} {\bibinfo {author} {\bibfnamefont {P.}~\bibnamefont
  {{Kocherlakota}}}\ and\ \bibinfo {author} {\bibfnamefont {L.}~\bibnamefont
  {{Rezzolla}}},\ }\bibfield  {title} {\bibinfo {title} {{Distinguishing
  gravitational and emission physics in black hole imaging: spherical
  symmetry}},\ }\href {https://doi.org/10.1093/mnras/stac891} {\bibfield
  {journal} {\bibinfo  {journal} {mnras}\ }\textbf {\bibinfo {volume} {513}},\
  \bibinfo {pages} {1229} (\bibinfo {year} {2022})},\ \Eprint
  {https://arxiv.org/abs/2201.05641} {arXiv:2201.05641 [gr-qc]} \BibitemShut
  {NoStop}%
\bibitem [{\citenamefont {{{\"O}zel}}\ \emph {et~al.}(2022)\citenamefont
  {{{\"O}zel}}, \citenamefont {{Psaltis}},\ and\ \citenamefont
  {{Younsi}}}]{Ozel+2021}%
  \BibitemOpen
  \bibfield  {author} {\bibinfo {author} {\bibfnamefont {F.}~\bibnamefont
  {{{\"O}zel}}}, \bibinfo {author} {\bibfnamefont {D.}~\bibnamefont
  {{Psaltis}}},\ and\ \bibinfo {author} {\bibfnamefont {Z.}~\bibnamefont
  {{Younsi}}},\ }\bibfield  {title} {\bibinfo {title} {{Black Hole Images as
  Tests of General Relativity: Effects of Plasma Physics}},\ }\href
  {https://doi.org/10.3847/1538-4357/ac9fcb} {\bibfield  {journal} {\bibinfo
  {journal} {\apj}\ }\textbf {\bibinfo {volume} {941}},\ \bibinfo {eid} {88}
  (\bibinfo {year} {2022})},\ \Eprint {https://arxiv.org/abs/2111.01123}
  {arXiv:2111.01123 [astro-ph.HE]} \BibitemShut {NoStop}%
\bibitem [{\citenamefont {{Younsi}}\ \emph {et~al.}(2023)\citenamefont
  {{Younsi}}, \citenamefont {{Psaltis}},\ and\ \citenamefont
  {{{\"O}zel}}}]{Younsi+2021}%
  \BibitemOpen
  \bibfield  {author} {\bibinfo {author} {\bibfnamefont {Z.}~\bibnamefont
  {{Younsi}}}, \bibinfo {author} {\bibfnamefont {D.}~\bibnamefont
  {{Psaltis}}},\ and\ \bibinfo {author} {\bibfnamefont {F.}~\bibnamefont
  {{{\"O}zel}}},\ }\bibfield  {title} {\bibinfo {title} {{Black Hole Images as
  Tests of General Relativity: Effects of Spacetime Geometry}},\ }\href
  {https://doi.org/10.3847/1538-4357/aca58a} {\bibfield  {journal} {\bibinfo
  {journal} {\apj}\ }\textbf {\bibinfo {volume} {942}},\ \bibinfo {eid} {47}
  (\bibinfo {year} {2023})},\ \Eprint {https://arxiv.org/abs/2111.01752}
  {arXiv:2111.01752 [astro-ph.HE]} \BibitemShut {NoStop}%
\bibitem [{\citenamefont {Akiyama}\ \emph
  {et~al.}(2022{\natexlab{b}})\citenamefont {Akiyama} \emph
  {et~al.}}]{EHTC+2022f}%
  \BibitemOpen
  \bibfield  {author} {\bibinfo {author} {\bibfnamefont {K.}~\bibnamefont
  {Akiyama}} \emph {et~al.} (\bibinfo {collaboration} {Event Horizon
  Telescope}),\ }\bibfield  {title} {\bibinfo {title} {{First Sagittarius A*
  Event Horizon Telescope Results. VI. Testing the Black Hole Metric}},\ }\href
  {https://doi.org/10.3847/2041-8213/ac6756} {\bibfield  {journal} {\bibinfo
  {journal} {Astrophys. J. Lett.}\ }\textbf {\bibinfo {volume} {930}},\
  \bibinfo {pages} {L17} (\bibinfo {year} {2022}{\natexlab{b}})},\ \Eprint
  {https://arxiv.org/abs/2311.09484} {arXiv:2311.09484 [astro-ph.HE]}
  \BibitemShut {NoStop}%
\bibitem [{\citenamefont {{Hioki}}\ and\ \citenamefont
  {{Maeda}}(2009)}]{Hioki2009}%
  \BibitemOpen
  \bibfield  {author} {\bibinfo {author} {\bibfnamefont {K.}~\bibnamefont
  {{Hioki}}}\ and\ \bibinfo {author} {\bibfnamefont {K.-I.}\ \bibnamefont
  {{Maeda}}},\ }\bibfield  {title} {\bibinfo {title} {{Measurement of the Kerr
  spin parameter by observation of a compact object's shadow}},\ }\href
  {https://doi.org/10.1103/PhysRevD.80.024042} {\bibfield  {journal} {\bibinfo
  {journal} {\prd}\ }\textbf {\bibinfo {volume} {80}},\ \bibinfo {eid} {024042}
  (\bibinfo {year} {2009})},\ \Eprint {https://arxiv.org/abs/0904.3575}
  {arXiv:0904.3575 [astro-ph.HE]} \BibitemShut {NoStop}%
\bibitem [{\citenamefont {Psaltis}\ \emph {et~al.}(2020)\citenamefont {Psaltis}
  \emph {et~al.}}]{Psaltis+2020}%
  \BibitemOpen
  \bibfield  {author} {\bibinfo {author} {\bibfnamefont {D.}~\bibnamefont
  {Psaltis}} \emph {et~al.} (\bibinfo {collaboration} {Event Horizon
  Telescope}),\ }\bibfield  {title} {\bibinfo {title} {{Gravitational Test
  Beyond the First Post-Newtonian Order with the Shadow of the M87 Black
  Hole}},\ }\href {https://doi.org/10.1103/PhysRevLett.125.141104} {\bibfield
  {journal} {\bibinfo  {journal} {Phys. Rev. Lett.}\ }\textbf {\bibinfo
  {volume} {125}},\ \bibinfo {pages} {141104} (\bibinfo {year} {2020})},\
  \Eprint {https://arxiv.org/abs/2010.01055} {arXiv:2010.01055 [gr-qc]}
  \BibitemShut {NoStop}%
\bibitem [{\citenamefont {{Bardeen}}(1973)}]{Bardeen1973}%
  \BibitemOpen
  \bibfield  {author} {\bibinfo {author} {\bibfnamefont {J.~M.}\ \bibnamefont
  {{Bardeen}}},\ }\bibfield  {title} {\bibinfo {title} {{Timelike and null
  geodesics in the Kerr metric.}},\ }in\ \href
  {https://adsabs.harvard.edu/full/1974IAUS...64..132B} {\emph {\bibinfo
  {booktitle} {Black Holes (Les Astres Occlus)}}}\ (\bibinfo {year} {1973})\
  pp.\ \bibinfo {pages} {215--239}\BibitemShut {NoStop}%
\bibitem [{\citenamefont {Luminet}(1979)}]{Luminet:1979nyg}%
  \BibitemOpen
  \bibfield  {author} {\bibinfo {author} {\bibfnamefont {J.~P.}\ \bibnamefont
  {Luminet}},\ }\bibfield  {title} {\bibinfo {title} {{Image of a spherical
  black hole with thin accretion disk}},\ }\href@noop {} {\bibfield  {journal}
  {\bibinfo  {journal} {Astron. Astrophys.}\ }\textbf {\bibinfo {volume}
  {75}},\ \bibinfo {pages} {228} (\bibinfo {year} {1979})}\BibitemShut
  {NoStop}%
\bibitem [{\citenamefont {{Gralla}}\ \emph {et~al.}(2019)\citenamefont
  {{Gralla}}, \citenamefont {{Holz}},\ and\ \citenamefont
  {{Wald}}}]{Gralla:2019xty}%
  \BibitemOpen
  \bibfield  {author} {\bibinfo {author} {\bibfnamefont {S.~E.}\ \bibnamefont
  {{Gralla}}}, \bibinfo {author} {\bibfnamefont {D.~E.}\ \bibnamefont
  {{Holz}}},\ and\ \bibinfo {author} {\bibfnamefont {R.~M.}\ \bibnamefont
  {{Wald}}},\ }\bibfield  {title} {\bibinfo {title} {{Black hole shadows,
  photon rings, and lensing rings}},\ }\href
  {https://doi.org/10.1103/PhysRevD.100.024018} {\bibfield  {journal} {\bibinfo
   {journal} {\prd}\ }\textbf {\bibinfo {volume} {100}},\ \bibinfo {eid}
  {024018} (\bibinfo {year} {2019})},\ \Eprint
  {https://arxiv.org/abs/1906.00873} {arXiv:1906.00873 [astro-ph.HE]}
  \BibitemShut {NoStop}%
\bibitem [{\citenamefont {{Johnson}}\ \emph {et~al.}(2020)\citenamefont
  {{Johnson}} \emph {et~al.}}]{Johnson:2019ljv}%
  \BibitemOpen
  \bibfield  {author} {\bibinfo {author} {\bibfnamefont {M.~D.}\ \bibnamefont
  {{Johnson}}} \emph {et~al.},\ }\bibfield  {title} {\bibinfo {title}
  {{Universal interferometric signatures of a black hole's photon ring}},\
  }\href {https://doi.org/10.1126/sciadv.aaz1310} {\bibfield  {journal}
  {\bibinfo  {journal} {sciadv}\ }\textbf {\bibinfo {volume} {6}},\ \bibinfo
  {pages} {eaaz1310} (\bibinfo {year} {2020})},\ \Eprint
  {https://arxiv.org/abs/1907.04329} {arXiv:1907.04329 [astro-ph.IM]}
  \BibitemShut {NoStop}%
\bibitem [{\citenamefont {{Gralla}}\ and\ \citenamefont
  {{Lupsasca}}(2020)}]{Gralla:2019drh}%
  \BibitemOpen
  \bibfield  {author} {\bibinfo {author} {\bibfnamefont {S.~E.}\ \bibnamefont
  {{Gralla}}}\ and\ \bibinfo {author} {\bibfnamefont {A.}~\bibnamefont
  {{Lupsasca}}},\ }\bibfield  {title} {\bibinfo {title} {{Lensing by Kerr black
  holes}},\ }\href {https://doi.org/10.1103/PhysRevD.101.044031} {\bibfield
  {journal} {\bibinfo  {journal} {\prd}\ }\textbf {\bibinfo {volume} {101}},\
  \bibinfo {eid} {044031} (\bibinfo {year} {2020})},\ \Eprint
  {https://arxiv.org/abs/1910.12873} {arXiv:1910.12873 [gr-qc]} \BibitemShut
  {NoStop}%
\bibitem [{\citenamefont {Gralla}\ and\ \citenamefont
  {Lupsasca}(2020)}]{Gralla:2020yvo}%
  \BibitemOpen
  \bibfield  {author} {\bibinfo {author} {\bibfnamefont {S.~E.}\ \bibnamefont
  {Gralla}}\ and\ \bibinfo {author} {\bibfnamefont {A.}~\bibnamefont
  {Lupsasca}},\ }\bibfield  {title} {\bibinfo {title} {{Observable shape of
  black hole photon rings}},\ }\href
  {https://doi.org/10.1103/PhysRevD.102.124003} {\bibfield  {journal} {\bibinfo
   {journal} {Phys. Rev. D}\ }\textbf {\bibinfo {volume} {102}},\ \bibinfo
  {pages} {124003} (\bibinfo {year} {2020})},\ \Eprint
  {https://arxiv.org/abs/2007.10336} {arXiv:2007.10336 [gr-qc]} \BibitemShut
  {NoStop}%
\bibitem [{\citenamefont {{Paugnat}}\ \emph {et~al.}(2022)\citenamefont
  {{Paugnat}}, \citenamefont {{Lupsasca}}, \citenamefont {{Vincent}},\ and\
  \citenamefont {{Wielgus}}}]{Paugnat+2022}%
  \BibitemOpen
  \bibfield  {author} {\bibinfo {author} {\bibfnamefont {H.}~\bibnamefont
  {{Paugnat}}}, \bibinfo {author} {\bibfnamefont {A.}~\bibnamefont
  {{Lupsasca}}}, \bibinfo {author} {\bibfnamefont {F.~H.}\ \bibnamefont
  {{Vincent}}},\ and\ \bibinfo {author} {\bibfnamefont {M.}~\bibnamefont
  {{Wielgus}}},\ }\bibfield  {title} {\bibinfo {title} {{Photon ring test of
  the Kerr hypothesis: Variation in the ring shape}},\ }\href
  {https://doi.org/10.1051/0004-6361/202244216} {\bibfield  {journal} {\bibinfo
   {journal} {aap}\ }\textbf {\bibinfo {volume} {668}},\ \bibinfo {eid} {A11}
  (\bibinfo {year} {2022})},\ \Eprint {https://arxiv.org/abs/2206.02781}
  {arXiv:2206.02781 [astro-ph.HE]} \BibitemShut {NoStop}%
\bibitem [{\citenamefont {C\'ardenas-Avenda\~no}\ and\ \citenamefont
  {Lupsasca}(2023)}]{Cardenas-Avendano:2023dzo}%
  \BibitemOpen
  \bibfield  {author} {\bibinfo {author} {\bibfnamefont {A.}~\bibnamefont
  {C\'ardenas-Avenda\~no}}\ and\ \bibinfo {author} {\bibfnamefont
  {A.}~\bibnamefont {Lupsasca}},\ }\bibfield  {title} {\bibinfo {title}
  {{Prediction for the interferometric shape of the first black hole photon
  ring}},\ }\href {https://doi.org/10.1103/PhysRevD.108.064043} {\bibfield
  {journal} {\bibinfo  {journal} {Phys. Rev. D}\ }\textbf {\bibinfo {volume}
  {108}},\ \bibinfo {pages} {064043} (\bibinfo {year} {2023})},\ \Eprint
  {https://arxiv.org/abs/2305.12956} {arXiv:2305.12956 [gr-qc]} \BibitemShut
  {NoStop}%
\bibitem [{\citenamefont {Palumbo}\ \emph {et~al.}(2023)\citenamefont
  {Palumbo}, \citenamefont {Wong}, \citenamefont {Chael},\ and\ \citenamefont
  {Johnson}}]{Palumbo:2023auc}%
  \BibitemOpen
  \bibfield  {author} {\bibinfo {author} {\bibfnamefont {D.~C.~M.}\
  \bibnamefont {Palumbo}}, \bibinfo {author} {\bibfnamefont {G.~N.}\
  \bibnamefont {Wong}}, \bibinfo {author} {\bibfnamefont {A.~A.}\ \bibnamefont
  {Chael}},\ and\ \bibinfo {author} {\bibfnamefont {M.~D.}\ \bibnamefont
  {Johnson}},\ }\bibfield  {title} {\bibinfo {title} {{Demonstrating Photon
  Ring Existence with Single-baseline Polarimetry}},\ }\href
  {https://doi.org/10.3847/2041-8213/ace630} {\bibfield  {journal} {\bibinfo
  {journal} {Astrophys. J. Lett.}\ }\textbf {\bibinfo {volume} {952}},\
  \bibinfo {pages} {L31} (\bibinfo {year} {2023})},\ \Eprint
  {https://arxiv.org/abs/2307.05293} {arXiv:2307.05293 [astro-ph.HE]}
  \BibitemShut {NoStop}%
\bibitem [{\citenamefont {{Chatterjee}}\ \emph
  {et~al.}(2023{\natexlab{a}})\citenamefont {{Chatterjee}}, \citenamefont
  {{Kocherlakota}}, \citenamefont {{Younsi}},\ and\ \citenamefont
  {{Narayan}}}]{Chatterjee+2023a}%
  \BibitemOpen
  \bibfield  {author} {\bibinfo {author} {\bibfnamefont {K.}~\bibnamefont
  {{Chatterjee}}}, \bibinfo {author} {\bibfnamefont {P.}~\bibnamefont
  {{Kocherlakota}}}, \bibinfo {author} {\bibfnamefont {Z.}~\bibnamefont
  {{Younsi}}},\ and\ \bibinfo {author} {\bibfnamefont {R.}~\bibnamefont
  {{Narayan}}},\ }\bibfield  {title} {\bibinfo {title} {{Energy Extraction from
  Spinning Stringy Black Holes}},\ }\href
  {https://doi.org/10.48550/arXiv.2310.20040} {\bibfield  {journal} {\bibinfo
  {journal} {arXiv e-prints}\ ,\ \bibinfo {eid} {arXiv:2310.20040}} (\bibinfo
  {year} {2023}{\natexlab{a}})},\ \Eprint {https://arxiv.org/abs/2310.20040}
  {arXiv:2310.20040 [gr-qc]} \BibitemShut {NoStop}%
\bibitem [{\citenamefont {{Chatterjee}}\ \emph
  {et~al.}(2023{\natexlab{b}})\citenamefont {{Chatterjee}}, \citenamefont
  {{Younsi}}, \citenamefont {{Kocherlakota}},\ and\ \citenamefont
  {{Narayan}}}]{Chatterjee+2023b}%
  \BibitemOpen
  \bibfield  {author} {\bibinfo {author} {\bibfnamefont {K.}~\bibnamefont
  {{Chatterjee}}}, \bibinfo {author} {\bibfnamefont {Z.}~\bibnamefont
  {{Younsi}}}, \bibinfo {author} {\bibfnamefont {P.}~\bibnamefont
  {{Kocherlakota}}},\ and\ \bibinfo {author} {\bibfnamefont {R.}~\bibnamefont
  {{Narayan}}},\ }\bibfield  {title} {\bibinfo {title} {{On the Universality of
  Energy Extraction from Black Hole Spacetimes}},\ }\href
  {https://doi.org/10.48550/arXiv.2310.20043} {\bibfield  {journal} {\bibinfo
  {journal} {arXiv e-prints}\ ,\ \bibinfo {eid} {arXiv:2310.20043}} (\bibinfo
  {year} {2023}{\natexlab{b}})},\ \Eprint {https://arxiv.org/abs/2310.20043}
  {arXiv:2310.20043 [gr-qc]} \BibitemShut {NoStop}%
\bibitem [{\citenamefont {{Johnson}}\ \emph {et~al.}(2023)\citenamefont
  {{Johnson}} \emph {et~al.}}]{Johnson+2023}%
  \BibitemOpen
  \bibfield  {author} {\bibinfo {author} {\bibfnamefont {M.~D.}\ \bibnamefont
  {{Johnson}}} \emph {et~al.},\ }\bibfield  {title} {\bibinfo {title} {{Key
  Science Goals for the Next-Generation Event Horizon Telescope}},\ }\href
  {https://doi.org/10.3390/galaxies11030061} {\bibfield  {journal} {\bibinfo
  {journal} {Galaxies}\ }\textbf {\bibinfo {volume} {11}},\ \bibinfo {eid} {61}
  (\bibinfo {year} {2023})}\BibitemShut {NoStop}%
\bibitem [{\citenamefont {Johnson}\ \emph {et~al.}(2024)\citenamefont {Johnson}
  \emph {et~al.}}]{Johnson:2024ttr}%
  \BibitemOpen
  \bibfield  {author} {\bibinfo {author} {\bibfnamefont {M.~D.}\ \bibnamefont
  {Johnson}} \emph {et~al.},\ }\bibfield  {title} {\bibinfo {title} {{The Black
  Hole Explorer: motivation and vision}},\ }\href
  {https://doi.org/10.1117/12.3019835} {\bibfield  {journal} {\bibinfo
  {journal} {Proc. SPIE Int. Soc. Opt. Eng.}\ }\textbf {\bibinfo {volume}
  {13092}},\ \bibinfo {pages} {130922D} (\bibinfo {year} {2024})},\ \Eprint
  {https://arxiv.org/abs/2406.12917} {arXiv:2406.12917 [astro-ph.IM]}
  \BibitemShut {NoStop}%
\bibitem [{\citenamefont {Galison}\ \emph {et~al.}(2024)\citenamefont
  {Galison}, \citenamefont {Johnson}, \citenamefont {Lupsasca}, \citenamefont
  {Gravely},\ and\ \citenamefont {Berens}}]{Galison:2024bop}%
  \BibitemOpen
  \bibfield  {author} {\bibinfo {author} {\bibfnamefont {P.}~\bibnamefont
  {Galison}}, \bibinfo {author} {\bibfnamefont {M.~D.}\ \bibnamefont
  {Johnson}}, \bibinfo {author} {\bibfnamefont {A.}~\bibnamefont {Lupsasca}},
  \bibinfo {author} {\bibfnamefont {T.}~\bibnamefont {Gravely}},\ and\ \bibinfo
  {author} {\bibfnamefont {R.}~\bibnamefont {Berens}},\ }\bibfield  {title}
  {\bibinfo {title} {{The Black Hole Explorer: using the photon ring to
  visualize spacetime around the black hole}},\ }\href
  {https://doi.org/10.1117/12.3019994} {\bibfield  {journal} {\bibinfo
  {journal} {Proc. SPIE Int. Soc. Opt. Eng.}\ }\textbf {\bibinfo {volume}
  {13092}},\ \bibinfo {pages} {130926R} (\bibinfo {year} {2024})},\ \Eprint
  {https://arxiv.org/abs/2406.11671} {arXiv:2406.11671 [gr-qc]} \BibitemShut
  {NoStop}%
\bibitem [{\citenamefont {Kawashima}\ \emph {et~al.}(2024)\citenamefont
  {Kawashima}, \citenamefont {Tsunetoe}, \citenamefont {Ohsuga}, \citenamefont
  {Kino}, \citenamefont {Mizuno}, \citenamefont {Moriyama}, \citenamefont
  {Saida}, \citenamefont {Akiyama}, \citenamefont {Hada},\ and\ \citenamefont
  {Niinuma}}]{Kawashima:2024svy}%
  \BibitemOpen
  \bibfield  {author} {\bibinfo {author} {\bibfnamefont {T.}~\bibnamefont
  {Kawashima}}, \bibinfo {author} {\bibfnamefont {Y.}~\bibnamefont {Tsunetoe}},
  \bibinfo {author} {\bibfnamefont {K.}~\bibnamefont {Ohsuga}}, \bibinfo
  {author} {\bibfnamefont {M.}~\bibnamefont {Kino}}, \bibinfo {author}
  {\bibfnamefont {Y.}~\bibnamefont {Mizuno}}, \bibinfo {author} {\bibfnamefont
  {K.}~\bibnamefont {Moriyama}}, \bibinfo {author} {\bibfnamefont
  {H.}~\bibnamefont {Saida}}, \bibinfo {author} {\bibfnamefont
  {K.}~\bibnamefont {Akiyama}}, \bibinfo {author} {\bibfnamefont
  {K.}~\bibnamefont {Hada}},\ and\ \bibinfo {author} {\bibfnamefont
  {K.}~\bibnamefont {Niinuma}},\ }\bibfield  {title} {\bibinfo {title} {{Black
  hole spacetime and properties of accretion flows and jets probed by Black
  Hole Explorer: science cases proposed by BHEX Japan team}},\ }\href
  {https://doi.org/10.1117/12.3018053} {\bibfield  {journal} {\bibinfo
  {journal} {Proc. SPIE Int. Soc. Opt. Eng.}\ }\textbf {\bibinfo {volume}
  {13092}},\ \bibinfo {pages} {130926X} (\bibinfo {year} {2024})},\ \Eprint
  {https://arxiv.org/abs/2406.09995} {arXiv:2406.09995 [astro-ph.HE]}
  \BibitemShut {NoStop}%
\bibitem [{\citenamefont {Lupsasca}\ \emph {et~al.}(2024)\citenamefont
  {Lupsasca}, \citenamefont {C\'ardenas-Avenda\~no}, \citenamefont {Palumbo},
  \citenamefont {Johnson}, \citenamefont {Gralla}, \citenamefont {Marrone},
  \citenamefont {Galison}, \citenamefont {Tiede},\ and\ \citenamefont
  {Keeble}}]{Lupsasca:2024xhq}%
  \BibitemOpen
  \bibfield  {author} {\bibinfo {author} {\bibfnamefont {A.}~\bibnamefont
  {Lupsasca}}, \bibinfo {author} {\bibfnamefont {A.}~\bibnamefont
  {C\'ardenas-Avenda\~no}}, \bibinfo {author} {\bibfnamefont {D.~C.~M.}\
  \bibnamefont {Palumbo}}, \bibinfo {author} {\bibfnamefont {M.~D.}\
  \bibnamefont {Johnson}}, \bibinfo {author} {\bibfnamefont {S.~E.}\
  \bibnamefont {Gralla}}, \bibinfo {author} {\bibfnamefont {D.~P.}\
  \bibnamefont {Marrone}}, \bibinfo {author} {\bibfnamefont {P.}~\bibnamefont
  {Galison}}, \bibinfo {author} {\bibfnamefont {P.}~\bibnamefont {Tiede}},\
  and\ \bibinfo {author} {\bibfnamefont {L.}~\bibnamefont {Keeble}},\
  }\bibfield  {title} {\bibinfo {title} {{The Black Hole Explorer: photon ring
  science, detection, and shape measurement}},\ }\href
  {https://doi.org/10.1117/12.3019437} {\bibfield  {journal} {\bibinfo
  {journal} {Proc. SPIE Int. Soc. Opt. Eng.}\ }\textbf {\bibinfo {volume}
  {13092}},\ \bibinfo {pages} {130926Q} (\bibinfo {year} {2024})},\ \Eprint
  {https://arxiv.org/abs/2406.09498} {arXiv:2406.09498 [gr-qc]} \BibitemShut
  {NoStop}%
\bibitem [{\citenamefont {Akiyama}\ \emph
  {et~al.}(2024{\natexlab{b}})\citenamefont {Akiyama} \emph
  {et~al.}}]{Akiyama:2024msp}%
  \BibitemOpen
  \bibfield  {author} {\bibinfo {author} {\bibfnamefont {K.}~\bibnamefont
  {Akiyama}} \emph {et~al.},\ }\bibfield  {title} {\bibinfo {title} {{The
  Japanese vision for the Black Hole Explorer mission}},\ }\href
  {https://doi.org/10.1117/12.3019968} {\bibfield  {journal} {\bibinfo
  {journal} {Proc. SPIE Int. Soc. Opt. Eng.}\ }\textbf {\bibinfo {volume}
  {13092}},\ \bibinfo {pages} {130922E} (\bibinfo {year}
  {2024}{\natexlab{b}})},\ \Eprint {https://arxiv.org/abs/2406.09516}
  {arXiv:2406.09516 [astro-ph.IM]} \BibitemShut {NoStop}%
\bibitem [{\citenamefont {Akiyama}\ \emph
  {et~al.}(2019{\natexlab{b}})\citenamefont {Akiyama} \emph
  {et~al.}}]{EHTC+2019f}%
  \BibitemOpen
  \bibfield  {author} {\bibinfo {author} {\bibfnamefont {K.}~\bibnamefont
  {Akiyama}} \emph {et~al.} (\bibinfo {collaboration} {Event Horizon
  Telescope}),\ }\bibfield  {title} {\bibinfo {title} {{First M87 Event Horizon
  Telescope Results. VI. The Shadow and Mass of the Central Black Hole}},\
  }\href {https://doi.org/10.3847/2041-8213/ab1141} {\bibfield  {journal}
  {\bibinfo  {journal} {Astrophys. J. Lett.}\ }\textbf {\bibinfo {volume}
  {875}},\ \bibinfo {pages} {L6} (\bibinfo {year} {2019}{\natexlab{b}})},\
  \Eprint {https://arxiv.org/abs/1906.11243} {arXiv:1906.11243 [astro-ph.GA]}
  \BibitemShut {NoStop}%
\bibitem [{\citenamefont {Kocherlakota}\ \emph {et~al.}(2021)\citenamefont
  {Kocherlakota} \emph {et~al.}}]{Kocherlakota+2021}%
  \BibitemOpen
  \bibfield  {author} {\bibinfo {author} {\bibfnamefont {P.}~\bibnamefont
  {Kocherlakota}} \emph {et~al.} (\bibinfo {collaboration} {Event Horizon
  Telescope}),\ }\bibfield  {title} {\bibinfo {title} {{Constraints on
  black-hole charges with the 2017 EHT observations of M87*}},\ }\href
  {https://doi.org/10.1103/PhysRevD.103.104047} {\bibfield  {journal} {\bibinfo
   {journal} {Phys. Rev. D}\ }\textbf {\bibinfo {volume} {103}},\ \bibinfo
  {pages} {104047} (\bibinfo {year} {2021})},\ \Eprint
  {https://arxiv.org/abs/2105.09343} {arXiv:2105.09343 [gr-qc]} \BibitemShut
  {NoStop}%
\bibitem [{\citenamefont {Kumar~Walia}\ \emph {et~al.}(2022)\citenamefont
  {Kumar~Walia}, \citenamefont {Ghosh},\ and\ \citenamefont
  {Maharaj}}]{KumarWalia:2022aop}%
  \BibitemOpen
  \bibfield  {author} {\bibinfo {author} {\bibfnamefont {R.}~\bibnamefont
  {Kumar~Walia}}, \bibinfo {author} {\bibfnamefont {S.~G.}\ \bibnamefont
  {Ghosh}},\ and\ \bibinfo {author} {\bibfnamefont {S.~D.}\ \bibnamefont
  {Maharaj}},\ }\bibfield  {title} {\bibinfo {title} {{Testing Rotating Regular
  Metrics with EHT Results of Sgr A*}},\ }\href
  {https://doi.org/10.3847/1538-4357/ac9623} {\bibfield  {journal} {\bibinfo
  {journal} {Astrophys. J.}\ }\textbf {\bibinfo {volume} {939}},\ \bibinfo
  {pages} {77} (\bibinfo {year} {2022})},\ \Eprint
  {https://arxiv.org/abs/2207.00078} {arXiv:2207.00078 [gr-qc]} \BibitemShut
  {NoStop}%
\bibitem [{\citenamefont {Vagnozzi}\ \emph {et~al.}(2023)\citenamefont
  {Vagnozzi} \emph {et~al.}}]{Vagnozzi:2022moj}%
  \BibitemOpen
  \bibfield  {author} {\bibinfo {author} {\bibfnamefont {S.}~\bibnamefont
  {Vagnozzi}} \emph {et~al.},\ }\bibfield  {title} {\bibinfo {title}
  {{Horizon-scale tests of gravity theories and fundamental physics from the
  Event Horizon Telescope image of Sagittarius A}},\ }\href
  {https://doi.org/10.1088/1361-6382/acd97b} {\bibfield  {journal} {\bibinfo
  {journal} {Class. Quant. Grav.}\ }\textbf {\bibinfo {volume} {40}},\ \bibinfo
  {pages} {165007} (\bibinfo {year} {2023})},\ \Eprint
  {https://arxiv.org/abs/2205.07787} {arXiv:2205.07787 [gr-qc]} \BibitemShut
  {NoStop}%
\bibitem [{\citenamefont {Kumar}\ \emph {et~al.}(2020)\citenamefont {Kumar},
  \citenamefont {Kumar},\ and\ \citenamefont {Ghosh}}]{Kumar:2020yem}%
  \BibitemOpen
  \bibfield  {author} {\bibinfo {author} {\bibfnamefont {R.}~\bibnamefont
  {Kumar}}, \bibinfo {author} {\bibfnamefont {A.}~\bibnamefont {Kumar}},\ and\
  \bibinfo {author} {\bibfnamefont {S.~G.}\ \bibnamefont {Ghosh}},\ }\bibfield
  {title} {\bibinfo {title} {{Testing Rotating Regular Metrics as Candidates
  for Astrophysical Black Holes}},\ }\href
  {https://doi.org/10.3847/1538-4357/ab8c4a} {\bibfield  {journal} {\bibinfo
  {journal} {Astrophys. J.}\ }\textbf {\bibinfo {volume} {896}},\ \bibinfo
  {pages} {89} (\bibinfo {year} {2020})},\ \Eprint
  {https://arxiv.org/abs/2006.09869} {arXiv:2006.09869 [gr-qc]} \BibitemShut
  {NoStop}%
\bibitem [{\citenamefont {Glampedakis}\ and\ \citenamefont
  {Pappas}(2023)}]{Glampedakis:2023eek}%
  \BibitemOpen
  \bibfield  {author} {\bibinfo {author} {\bibfnamefont {K.}~\bibnamefont
  {Glampedakis}}\ and\ \bibinfo {author} {\bibfnamefont {G.}~\bibnamefont
  {Pappas}},\ }\bibfield  {title} {\bibinfo {title} {{Is a black hole shadow a
  reliable test of the no-hair theorem?}},\ }\href
  {https://doi.org/10.1103/PhysRevD.107.064001} {\bibfield  {journal} {\bibinfo
   {journal} {Phys. Rev. D}\ }\textbf {\bibinfo {volume} {107}},\ \bibinfo
  {pages} {064001} (\bibinfo {year} {2023})},\ \Eprint
  {https://arxiv.org/abs/2302.06140} {arXiv:2302.06140 [gr-qc]} \BibitemShut
  {NoStop}%
\bibitem [{\citenamefont {Kumar}\ and\ \citenamefont
  {Ghosh}(2020)}]{Kumar:2018ple}%
  \BibitemOpen
  \bibfield  {author} {\bibinfo {author} {\bibfnamefont {R.}~\bibnamefont
  {Kumar}}\ and\ \bibinfo {author} {\bibfnamefont {S.~G.}\ \bibnamefont
  {Ghosh}},\ }\bibfield  {title} {\bibinfo {title} {{Black Hole Parameter
  Estimation from Its Shadow}},\ }\href
  {https://doi.org/10.3847/1538-4357/ab77b0} {\bibfield  {journal} {\bibinfo
  {journal} {Astrophys. J.}\ }\textbf {\bibinfo {volume} {892}},\ \bibinfo
  {pages} {78} (\bibinfo {year} {2020})},\ \Eprint
  {https://arxiv.org/abs/1811.01260} {arXiv:1811.01260 [gr-qc]} \BibitemShut
  {NoStop}%
\bibitem [{\citenamefont {Tsukamoto}\ \emph {et~al.}(2014)\citenamefont
  {Tsukamoto}, \citenamefont {Li},\ and\ \citenamefont
  {Bambi}}]{Tsukamoto:2014tja}%
  \BibitemOpen
  \bibfield  {author} {\bibinfo {author} {\bibfnamefont {N.}~\bibnamefont
  {Tsukamoto}}, \bibinfo {author} {\bibfnamefont {Z.}~\bibnamefont {Li}},\ and\
  \bibinfo {author} {\bibfnamefont {C.}~\bibnamefont {Bambi}},\ }\bibfield
  {title} {\bibinfo {title} {{Constraining the spin and the deformation
  parameters from the black hole shadow}},\ }\href
  {https://doi.org/10.1088/1475-7516/2014/06/043} {\bibfield  {journal}
  {\bibinfo  {journal} {JCAP}\ }\textbf {\bibinfo {volume} {06}},\ \bibinfo
  {pages} {043}},\ \Eprint {https://arxiv.org/abs/1403.0371} {arXiv:1403.0371
  [gr-qc]} \BibitemShut {NoStop}%
\bibitem [{\citenamefont {Gralla}(2020)}]{Gralla:2020nwp}%
  \BibitemOpen
  \bibfield  {author} {\bibinfo {author} {\bibfnamefont {S.~E.}\ \bibnamefont
  {Gralla}},\ }\bibfield  {title} {\bibinfo {title} {{Measuring the shape of a
  black hole photon ring}},\ }\href
  {https://doi.org/10.1103/PhysRevD.102.044017} {\bibfield  {journal} {\bibinfo
   {journal} {Phys. Rev. D}\ }\textbf {\bibinfo {volume} {102}},\ \bibinfo
  {pages} {044017} (\bibinfo {year} {2020})},\ \Eprint
  {https://arxiv.org/abs/2005.03856} {arXiv:2005.03856 [astro-ph.HE]}
  \BibitemShut {NoStop}%
\bibitem [{\citenamefont {{Vincent}}\ \emph {et~al.}(2022)\citenamefont
  {{Vincent}}, \citenamefont {{Gralla}}, \citenamefont {{Lupsasca}},\ and\
  \citenamefont {{Wielgus}}}]{Vincent+2022}%
  \BibitemOpen
  \bibfield  {author} {\bibinfo {author} {\bibfnamefont {F.~H.}\ \bibnamefont
  {{Vincent}}}, \bibinfo {author} {\bibfnamefont {S.~E.}\ \bibnamefont
  {{Gralla}}}, \bibinfo {author} {\bibfnamefont {A.}~\bibnamefont
  {{Lupsasca}}},\ and\ \bibinfo {author} {\bibfnamefont {M.}~\bibnamefont
  {{Wielgus}}},\ }\bibfield  {title} {\bibinfo {title} {{Images and photon ring
  signatures of thick disks around black holes}},\ }\href
  {https://doi.org/10.1051/0004-6361/202244339} {\bibfield  {journal} {\bibinfo
   {journal} {aap}\ }\textbf {\bibinfo {volume} {667}},\ \bibinfo {eid} {A170}
  (\bibinfo {year} {2022})},\ \Eprint {https://arxiv.org/abs/2206.12066}
  {arXiv:2206.12066 [astro-ph.HE]} \BibitemShut {NoStop}%
\bibitem [{\citenamefont {Kocherlakota}\ \emph {et~al.}(2024)\citenamefont
  {Kocherlakota}, \citenamefont {Rezzolla}, \citenamefont {Roy},\ and\
  \citenamefont {Wielgus}}]{Kocherlakota+2024a}%
  \BibitemOpen
  \bibfield  {author} {\bibinfo {author} {\bibfnamefont {P.}~\bibnamefont
  {Kocherlakota}}, \bibinfo {author} {\bibfnamefont {L.}~\bibnamefont
  {Rezzolla}}, \bibinfo {author} {\bibfnamefont {R.}~\bibnamefont {Roy}},\ and\
  \bibinfo {author} {\bibfnamefont {M.}~\bibnamefont {Wielgus}},\ }\bibfield
  {title} {\bibinfo {title} {{Prospects for future experimental tests of
  gravity with black hole imaging: Spherical symmetry}},\ }\href
  {https://doi.org/10.1103/PhysRevD.109.064064} {\bibfield  {journal} {\bibinfo
   {journal} {Phys. Rev. D}\ }\textbf {\bibinfo {volume} {109}},\ \bibinfo
  {pages} {064064} (\bibinfo {year} {2024})},\ \Eprint
  {https://arxiv.org/abs/2307.16841} {arXiv:2307.16841 [gr-qc]} \BibitemShut
  {NoStop}%
\bibitem [{\citenamefont {{Kocherlakota}}\ \emph {et~al.}(2024)\citenamefont
  {{Kocherlakota}}, \citenamefont {{Rezzolla}}, \citenamefont {{Roy}},\ and\
  \citenamefont {{Wielgus}}}]{Kocherlakota+2024b}%
  \BibitemOpen
  \bibfield  {author} {\bibinfo {author} {\bibfnamefont {P.}~\bibnamefont
  {{Kocherlakota}}}, \bibinfo {author} {\bibfnamefont {L.}~\bibnamefont
  {{Rezzolla}}}, \bibinfo {author} {\bibfnamefont {R.}~\bibnamefont {{Roy}}},\
  and\ \bibinfo {author} {\bibfnamefont {M.}~\bibnamefont {{Wielgus}}},\
  }\bibfield  {title} {\bibinfo {title} {{Hotspots and photon rings in
  spherically-symmetric spacetimes}},\ }\bibfield  {journal} {\bibinfo
  {journal} {mnras}\ }\href {https://doi.org/10.1093/mnras/stae1321}
  {10.1093/mnras/stae1321} (\bibinfo {year} {2024}),\ \Eprint
  {https://arxiv.org/abs/2403.08862} {arXiv:2403.08862 [astro-ph.HE]}
  \BibitemShut {NoStop}%
\bibitem [{\citenamefont {{Hadar}}\ \emph {et~al.}(2021)\citenamefont
  {{Hadar}}, \citenamefont {{Johnson}}, \citenamefont {{Lupsasca}},\ and\
  \citenamefont {{Wong}}}]{Hadar+2021}%
  \BibitemOpen
  \bibfield  {author} {\bibinfo {author} {\bibfnamefont {S.}~\bibnamefont
  {{Hadar}}}, \bibinfo {author} {\bibfnamefont {M.~D.}\ \bibnamefont
  {{Johnson}}}, \bibinfo {author} {\bibfnamefont {A.}~\bibnamefont
  {{Lupsasca}}},\ and\ \bibinfo {author} {\bibfnamefont {G.~N.}\ \bibnamefont
  {{Wong}}},\ }\bibfield  {title} {\bibinfo {title} {{Photon ring
  autocorrelations}},\ }\href {https://doi.org/10.1103/PhysRevD.103.104038}
  {\bibfield  {journal} {\bibinfo  {journal} {\prd}\ }\textbf {\bibinfo
  {volume} {103}},\ \bibinfo {eid} {104038} (\bibinfo {year} {2021})},\ \Eprint
  {https://arxiv.org/abs/2010.03683} {arXiv:2010.03683 [gr-qc]} \BibitemShut
  {NoStop}%
\bibitem [{\citenamefont {Hadar}\ \emph {et~al.}(2022)\citenamefont {Hadar},
  \citenamefont {Kapec}, \citenamefont {Lupsasca},\ and\ \citenamefont
  {Strominger}}]{Hadar+2022}%
  \BibitemOpen
  \bibfield  {author} {\bibinfo {author} {\bibfnamefont {S.}~\bibnamefont
  {Hadar}}, \bibinfo {author} {\bibfnamefont {D.}~\bibnamefont {Kapec}},
  \bibinfo {author} {\bibfnamefont {A.}~\bibnamefont {Lupsasca}},\ and\
  \bibinfo {author} {\bibfnamefont {A.}~\bibnamefont {Strominger}},\ }\bibfield
   {title} {\bibinfo {title} {{Holography of the photon ring}},\ }\href
  {https://doi.org/10.1088/1361-6382/ac8d43} {\bibfield  {journal} {\bibinfo
  {journal} {Class. Quant. Grav.}\ }\textbf {\bibinfo {volume} {39}},\ \bibinfo
  {pages} {215001} (\bibinfo {year} {2022})},\ \Eprint
  {https://arxiv.org/abs/2205.05064} {arXiv:2205.05064 [gr-qc]} \BibitemShut
  {NoStop}%
\bibitem [{\citenamefont {Himwich}\ \emph {et~al.}(2020)\citenamefont
  {Himwich}, \citenamefont {Johnson}, \citenamefont {Lupsasca},\ and\
  \citenamefont {Strominger}}]{Himwich:2020msm}%
  \BibitemOpen
  \bibfield  {author} {\bibinfo {author} {\bibfnamefont {E.}~\bibnamefont
  {Himwich}}, \bibinfo {author} {\bibfnamefont {M.~D.}\ \bibnamefont
  {Johnson}}, \bibinfo {author} {\bibfnamefont {A.}~\bibnamefont {Lupsasca}},\
  and\ \bibinfo {author} {\bibfnamefont {A.}~\bibnamefont {Strominger}},\
  }\bibfield  {title} {\bibinfo {title} {{Universal polarimetric signatures of
  the black hole photon ring}},\ }\href
  {https://doi.org/10.1103/PhysRevD.101.084020} {\bibfield  {journal} {\bibinfo
   {journal} {Phys. Rev. D}\ }\textbf {\bibinfo {volume} {101}},\ \bibinfo
  {pages} {084020} (\bibinfo {year} {2020})},\ \Eprint
  {https://arxiv.org/abs/2001.08750} {arXiv:2001.08750 [gr-qc]} \BibitemShut
  {NoStop}%
\bibitem [{\citenamefont {Jia}\ \emph {et~al.}(2024)\citenamefont {Jia},
  \citenamefont {Quataert}, \citenamefont {Lupsasca},\ and\ \citenamefont
  {Wong}}]{Jia:2024mlb}%
  \BibitemOpen
  \bibfield  {author} {\bibinfo {author} {\bibfnamefont {H.}~\bibnamefont
  {Jia}}, \bibinfo {author} {\bibfnamefont {E.}~\bibnamefont {Quataert}},
  \bibinfo {author} {\bibfnamefont {A.}~\bibnamefont {Lupsasca}},\ and\
  \bibinfo {author} {\bibfnamefont {G.~N.}\ \bibnamefont {Wong}},\ }\bibfield
  {title} {\bibinfo {title} {{Photon ring interferometric signatures beyond the
  universal regime}},\ }\href {https://doi.org/10.1103/PhysRevD.110.083044}
  {\bibfield  {journal} {\bibinfo  {journal} {Phys. Rev. D}\ }\textbf {\bibinfo
  {volume} {110}},\ \bibinfo {pages} {083044} (\bibinfo {year} {2024})},\
  \Eprint {https://arxiv.org/abs/2405.08804} {arXiv:2405.08804 [astro-ph.HE]}
  \BibitemShut {NoStop}%
\bibitem [{\citenamefont {Staelens}\ \emph {et~al.}(2023)\citenamefont
  {Staelens}, \citenamefont {Mayerson}, \citenamefont {Bacchini}, \citenamefont
  {Ripperda},\ and\ \citenamefont {K\"uchler}}]{Staelens:2023jgr}%
  \BibitemOpen
  \bibfield  {author} {\bibinfo {author} {\bibfnamefont {S.}~\bibnamefont
  {Staelens}}, \bibinfo {author} {\bibfnamefont {D.~R.}\ \bibnamefont
  {Mayerson}}, \bibinfo {author} {\bibfnamefont {F.}~\bibnamefont {Bacchini}},
  \bibinfo {author} {\bibfnamefont {B.}~\bibnamefont {Ripperda}},\ and\
  \bibinfo {author} {\bibfnamefont {L.}~\bibnamefont {K\"uchler}},\ }\bibfield
  {title} {\bibinfo {title} {{Black hole photon rings beyond general
  relativity}},\ }\href {https://doi.org/10.1103/PhysRevD.107.124026}
  {\bibfield  {journal} {\bibinfo  {journal} {Phys. Rev. D}\ }\textbf {\bibinfo
  {volume} {107}},\ \bibinfo {pages} {124026} (\bibinfo {year} {2023})},\
  \Eprint {https://arxiv.org/abs/2303.02111} {arXiv:2303.02111 [gr-qc]}
  \BibitemShut {NoStop}%
\bibitem [{\citenamefont {Wielgus}(2021)}]{Wielgus:2021peu}%
  \BibitemOpen
  \bibfield  {author} {\bibinfo {author} {\bibfnamefont {M.}~\bibnamefont
  {Wielgus}},\ }\bibfield  {title} {\bibinfo {title} {{Photon rings of
  spherically symmetric black holes and robust tests of non-Kerr metrics}},\
  }\href {https://doi.org/10.1103/PhysRevD.104.124058} {\bibfield  {journal}
  {\bibinfo  {journal} {Phys. Rev. D}\ }\textbf {\bibinfo {volume} {104}},\
  \bibinfo {pages} {124058} (\bibinfo {year} {2021})},\ \Eprint
  {https://arxiv.org/abs/2109.10840} {arXiv:2109.10840 [gr-qc]} \BibitemShut
  {NoStop}%
\bibitem [{\citenamefont {da~Silva}\ \emph {et~al.}(2023)\citenamefont
  {da~Silva}, \citenamefont {Lobo}, \citenamefont {Olmo},\ and\ \citenamefont
  {Rubiera-Garcia}}]{daSilva:2023jxa}%
  \BibitemOpen
  \bibfield  {author} {\bibinfo {author} {\bibfnamefont {L.~F.~D.}\
  \bibnamefont {da~Silva}}, \bibinfo {author} {\bibfnamefont {F.~S.~N.}\
  \bibnamefont {Lobo}}, \bibinfo {author} {\bibfnamefont {G.~J.}\ \bibnamefont
  {Olmo}},\ and\ \bibinfo {author} {\bibfnamefont {D.}~\bibnamefont
  {Rubiera-Garcia}},\ }\bibfield  {title} {\bibinfo {title} {{Photon rings as
  tests for alternative spherically symmetric geometries with thin accretion
  disks}},\ }\href {https://doi.org/10.1103/PhysRevD.108.084055} {\bibfield
  {journal} {\bibinfo  {journal} {Phys. Rev. D}\ }\textbf {\bibinfo {volume}
  {108}},\ \bibinfo {pages} {084055} (\bibinfo {year} {2023})},\ \Eprint
  {https://arxiv.org/abs/2307.06778} {arXiv:2307.06778 [gr-qc]} \BibitemShut
  {NoStop}%
\bibitem [{\citenamefont {Cunha}\ \emph {et~al.}(2017)\citenamefont {Cunha},
  \citenamefont {Berti},\ and\ \citenamefont {Herdeiro}}]{Cunha:2017qtt}%
  \BibitemOpen
  \bibfield  {author} {\bibinfo {author} {\bibfnamefont {P.~V.~P.}\
  \bibnamefont {Cunha}}, \bibinfo {author} {\bibfnamefont {E.}~\bibnamefont
  {Berti}},\ and\ \bibinfo {author} {\bibfnamefont {C.~A.~R.}\ \bibnamefont
  {Herdeiro}},\ }\bibfield  {title} {\bibinfo {title} {{Light-Ring Stability
  for Ultracompact Objects}},\ }\href
  {https://doi.org/10.1103/PhysRevLett.119.251102} {\bibfield  {journal}
  {\bibinfo  {journal} {Phys. Rev. Lett.}\ }\textbf {\bibinfo {volume} {119}},\
  \bibinfo {pages} {251102} (\bibinfo {year} {2017})},\ \Eprint
  {https://arxiv.org/abs/1708.04211} {arXiv:1708.04211 [gr-qc]} \BibitemShut
  {NoStop}%
\bibitem [{\citenamefont {Medeiros}\ \emph {et~al.}(2020)\citenamefont
  {Medeiros}, \citenamefont {Psaltis},\ and\ \citenamefont
  {\"Ozel}}]{Medeiros:2019cde}%
  \BibitemOpen
  \bibfield  {author} {\bibinfo {author} {\bibfnamefont {L.}~\bibnamefont
  {Medeiros}}, \bibinfo {author} {\bibfnamefont {D.}~\bibnamefont {Psaltis}},\
  and\ \bibinfo {author} {\bibfnamefont {F.}~\bibnamefont {\"Ozel}},\
  }\bibfield  {title} {\bibinfo {title} {{A Parametric model for the shapes of
  black-hole shadows in non-Kerr spacetimes}},\ }\href
  {https://doi.org/10.3847/1538-4357/ab8bd1} {\bibfield  {journal} {\bibinfo
  {journal} {Astrophys. J.}\ }\textbf {\bibinfo {volume} {896}},\ \bibinfo
  {pages} {7} (\bibinfo {year} {2020})},\ \Eprint
  {https://arxiv.org/abs/1907.12575} {arXiv:1907.12575 [astro-ph.HE]}
  \BibitemShut {NoStop}%
\bibitem [{\citenamefont {Salehi}\ \emph {et~al.}(2024)\citenamefont {Salehi},
  \citenamefont {Kumar~Walia}, \citenamefont {Chang},\ and\ \citenamefont
  {Kocherlakota}}]{Salehi:2024cim}%
  \BibitemOpen
  \bibfield  {author} {\bibinfo {author} {\bibfnamefont {K.}~\bibnamefont
  {Salehi}}, \bibinfo {author} {\bibfnamefont {R.}~\bibnamefont {Kumar~Walia}},
  \bibinfo {author} {\bibfnamefont {D.}~\bibnamefont {Chang}},\ and\ \bibinfo
  {author} {\bibfnamefont {P.}~\bibnamefont {Kocherlakota}},\ }\bibfield
  {title} {\bibinfo {title} {{Influence of Observer's Inclination and Spacetime
  Structure on Photon Ring Observables}},\ }\href@noop {} {\  (\bibinfo {year}
  {2024})},\ \Eprint {https://arxiv.org/abs/2411.15310} {arXiv:2411.15310
  [gr-qc]} \BibitemShut {NoStop}%
\bibitem [{\citenamefont {Johannsen}(2013{\natexlab{a}})}]{Johannsen:2013szh}%
  \BibitemOpen
  \bibfield  {author} {\bibinfo {author} {\bibfnamefont {T.}~\bibnamefont
  {Johannsen}},\ }\bibfield  {title} {\bibinfo {title} {{Regular Black Hole
  Metric with Three Constants of Motion}},\ }\href
  {https://doi.org/10.1103/PhysRevD.88.044002} {\bibfield  {journal} {\bibinfo
  {journal} {Phys. Rev. D}\ }\textbf {\bibinfo {volume} {88}},\ \bibinfo
  {pages} {044002} (\bibinfo {year} {2013}{\natexlab{a}})},\ \Eprint
  {https://arxiv.org/abs/1501.02809} {arXiv:1501.02809 [gr-qc]} \BibitemShut
  {NoStop}%
\bibitem [{\citenamefont {Salehi}\ \emph {et~al.}(2023)\citenamefont {Salehi},
  \citenamefont {Broderick},\ and\ \citenamefont {Georgiev}}]{Salehi+2023}%
  \BibitemOpen
  \bibfield  {author} {\bibinfo {author} {\bibfnamefont {K.}~\bibnamefont
  {Salehi}}, \bibinfo {author} {\bibfnamefont {A.}~\bibnamefont {Broderick}},\
  and\ \bibinfo {author} {\bibfnamefont {B.}~\bibnamefont {Georgiev}},\
  }\bibfield  {title} {\bibinfo {title} {{Photon Rings and Shadow Size for
  General Integrable Spacetimes}},\ }\href@noop {} {\  (\bibinfo {year}
  {2023})},\ \Eprint {https://arxiv.org/abs/2311.01495} {arXiv:2311.01495
  [gr-qc]} \BibitemShut {NoStop}%
\bibitem [{\citenamefont {Newman}\ \emph {et~al.}(1965)\citenamefont {Newman},
  \citenamefont {Couch}, \citenamefont {Chinnapared}, \citenamefont {Exton},
  \citenamefont {Prakash},\ and\ \citenamefont {Torrence}}]{Newman:1965my}%
  \BibitemOpen
  \bibfield  {author} {\bibinfo {author} {\bibfnamefont {E.~T.}\ \bibnamefont
  {Newman}}, \bibinfo {author} {\bibfnamefont {R.}~\bibnamefont {Couch}},
  \bibinfo {author} {\bibfnamefont {K.}~\bibnamefont {Chinnapared}}, \bibinfo
  {author} {\bibfnamefont {A.}~\bibnamefont {Exton}}, \bibinfo {author}
  {\bibfnamefont {A.}~\bibnamefont {Prakash}},\ and\ \bibinfo {author}
  {\bibfnamefont {R.}~\bibnamefont {Torrence}},\ }\bibfield  {title} {\bibinfo
  {title} {{Metric of a Rotating, Charged Mass}},\ }\href
  {https://doi.org/10.1063/1.1704351} {\bibfield  {journal} {\bibinfo
  {journal} {J. Math. Phys.}\ }\textbf {\bibinfo {volume} {6}},\ \bibinfo
  {pages} {918} (\bibinfo {year} {1965})}\BibitemShut {NoStop}%
\bibitem [{\citenamefont {Sen}(1992)}]{Sen:1992ua}%
  \BibitemOpen
  \bibfield  {author} {\bibinfo {author} {\bibfnamefont {A.}~\bibnamefont
  {Sen}},\ }\bibfield  {title} {\bibinfo {title} {{Rotating charged black hole
  solution in heterotic string theory}},\ }\href
  {https://doi.org/10.1103/PhysRevLett.69.1006} {\bibfield  {journal} {\bibinfo
   {journal} {Phys. Rev. Lett.}\ }\textbf {\bibinfo {volume} {69}},\ \bibinfo
  {pages} {1006} (\bibinfo {year} {1992})},\ \Eprint
  {https://arxiv.org/abs/hep-th/9204046} {arXiv:hep-th/9204046} \BibitemShut
  {NoStop}%
\bibitem [{\citenamefont {Bambi}\ and\ \citenamefont
  {Modesto}(2013)}]{Bambi:2013ufa}%
  \BibitemOpen
  \bibfield  {author} {\bibinfo {author} {\bibfnamefont {C.}~\bibnamefont
  {Bambi}}\ and\ \bibinfo {author} {\bibfnamefont {L.}~\bibnamefont
  {Modesto}},\ }\bibfield  {title} {\bibinfo {title} {{Rotating regular black
  holes}},\ }\href {https://doi.org/10.1016/j.physletb.2013.03.025} {\bibfield
  {journal} {\bibinfo  {journal} {Phys. Lett. B}\ }\textbf {\bibinfo {volume}
  {721}},\ \bibinfo {pages} {329} (\bibinfo {year} {2013})},\ \Eprint
  {https://arxiv.org/abs/1302.6075} {arXiv:1302.6075 [gr-qc]} \BibitemShut
  {NoStop}%
\bibitem [{\citenamefont {Younsi}\ \emph {et~al.}(2016)\citenamefont {Younsi},
  \citenamefont {Zhidenko}, \citenamefont {Rezzolla}, \citenamefont
  {Konoplya},\ and\ \citenamefont {Mizuno}}]{Younsi:2016azx}%
  \BibitemOpen
  \bibfield  {author} {\bibinfo {author} {\bibfnamefont {Z.}~\bibnamefont
  {Younsi}}, \bibinfo {author} {\bibfnamefont {A.}~\bibnamefont {Zhidenko}},
  \bibinfo {author} {\bibfnamefont {L.}~\bibnamefont {Rezzolla}}, \bibinfo
  {author} {\bibfnamefont {R.}~\bibnamefont {Konoplya}},\ and\ \bibinfo
  {author} {\bibfnamefont {Y.}~\bibnamefont {Mizuno}},\ }\bibfield  {title}
  {\bibinfo {title} {{New method for shadow calculations: Application to
  parametrized axisymmetric black holes}},\ }\href
  {https://doi.org/10.1103/PhysRevD.94.084025} {\bibfield  {journal} {\bibinfo
  {journal} {Phys. Rev. D}\ }\textbf {\bibinfo {volume} {94}},\ \bibinfo
  {pages} {084025} (\bibinfo {year} {2016})},\ \Eprint
  {https://arxiv.org/abs/1607.05767} {arXiv:1607.05767 [gr-qc]} \BibitemShut
  {NoStop}%
\bibitem [{\citenamefont {Konoplya}\ and\ \citenamefont
  {Zhidenko}(2021)}]{Konoplya:2021slg}%
  \BibitemOpen
  \bibfield  {author} {\bibinfo {author} {\bibfnamefont {R.~A.}\ \bibnamefont
  {Konoplya}}\ and\ \bibinfo {author} {\bibfnamefont {A.}~\bibnamefont
  {Zhidenko}},\ }\bibfield  {title} {\bibinfo {title} {{Shadows of parametrized
  axially symmetric black holes allowing for separation of variables}},\ }\href
  {https://doi.org/10.1103/PhysRevD.103.104033} {\bibfield  {journal} {\bibinfo
   {journal} {Phys. Rev. D}\ }\textbf {\bibinfo {volume} {103}},\ \bibinfo
  {pages} {104033} (\bibinfo {year} {2021})},\ \Eprint
  {https://arxiv.org/abs/2103.03855} {arXiv:2103.03855 [gr-qc]} \BibitemShut
  {NoStop}%
\bibitem [{\citenamefont {Younsi}\ \emph {et~al.}(2023)\citenamefont {Younsi},
  \citenamefont {Psaltis},\ and\ \citenamefont {\"Ozel}}]{Younsi:2021dxe}%
  \BibitemOpen
  \bibfield  {author} {\bibinfo {author} {\bibfnamefont {Z.}~\bibnamefont
  {Younsi}}, \bibinfo {author} {\bibfnamefont {D.}~\bibnamefont {Psaltis}},\
  and\ \bibinfo {author} {\bibfnamefont {F.}~\bibnamefont {\"Ozel}},\
  }\bibfield  {title} {\bibinfo {title} {{Black Hole Images as Tests of General
  Relativity: Effects of Spacetime Geometry}},\ }\href
  {https://doi.org/10.3847/1538-4357/aca58a} {\bibfield  {journal} {\bibinfo
  {journal} {Astrophys. J.}\ }\textbf {\bibinfo {volume} {942}},\ \bibinfo
  {pages} {47} (\bibinfo {year} {2023})},\ \Eprint
  {https://arxiv.org/abs/2111.01752} {arXiv:2111.01752 [astro-ph.HE]}
  \BibitemShut {NoStop}%
\bibitem [{\citenamefont {Johannsen}(2013{\natexlab{b}})}]{Johannsen:2013vgc}%
  \BibitemOpen
  \bibfield  {author} {\bibinfo {author} {\bibfnamefont {T.}~\bibnamefont
  {Johannsen}},\ }\bibfield  {title} {\bibinfo {title} {{Photon Rings around
  Kerr and Kerr-like Black Holes}},\ }\href
  {https://doi.org/10.1088/0004-637X/777/2/170} {\bibfield  {journal} {\bibinfo
   {journal} {Astrophys. J.}\ }\textbf {\bibinfo {volume} {777}},\ \bibinfo
  {pages} {170} (\bibinfo {year} {2013}{\natexlab{b}})},\ \Eprint
  {https://arxiv.org/abs/1501.02814} {arXiv:1501.02814 [astro-ph.HE]}
  \BibitemShut {NoStop}%
\bibitem [{\citenamefont {Glampedakis}\ and\ \citenamefont
  {Pappas}(2019)}]{Glampedakis:2018blj}%
  \BibitemOpen
  \bibfield  {author} {\bibinfo {author} {\bibfnamefont {K.}~\bibnamefont
  {Glampedakis}}\ and\ \bibinfo {author} {\bibfnamefont {G.}~\bibnamefont
  {Pappas}},\ }\bibfield  {title} {\bibinfo {title} {{Modification of photon
  trapping orbits as a diagnostic of non-Kerr spacetimes}},\ }\href
  {https://doi.org/10.1103/PhysRevD.99.124041} {\bibfield  {journal} {\bibinfo
  {journal} {Phys. Rev. D}\ }\textbf {\bibinfo {volume} {99}},\ \bibinfo
  {pages} {124041} (\bibinfo {year} {2019})},\ \Eprint
  {https://arxiv.org/abs/1806.09333} {arXiv:1806.09333 [gr-qc]} \BibitemShut
  {NoStop}%
\bibitem [{\citenamefont {Cunha}\ \emph {et~al.}(2016)\citenamefont {Cunha},
  \citenamefont {Grover}, \citenamefont {Herdeiro}, \citenamefont {Radu},
  \citenamefont {Runarsson},\ and\ \citenamefont {Wittig}}]{Cunha:2016bjh}%
  \BibitemOpen
  \bibfield  {author} {\bibinfo {author} {\bibfnamefont {P.~V.~P.}\
  \bibnamefont {Cunha}}, \bibinfo {author} {\bibfnamefont {J.}~\bibnamefont
  {Grover}}, \bibinfo {author} {\bibfnamefont {C.}~\bibnamefont {Herdeiro}},
  \bibinfo {author} {\bibfnamefont {E.}~\bibnamefont {Radu}}, \bibinfo {author}
  {\bibfnamefont {H.}~\bibnamefont {Runarsson}},\ and\ \bibinfo {author}
  {\bibfnamefont {A.}~\bibnamefont {Wittig}},\ }\bibfield  {title} {\bibinfo
  {title} {{Chaotic lensing around boson stars and Kerr black holes with scalar
  hair}},\ }\href {https://doi.org/10.1103/PhysRevD.94.104023} {\bibfield
  {journal} {\bibinfo  {journal} {Phys. Rev. D}\ }\textbf {\bibinfo {volume}
  {94}},\ \bibinfo {pages} {104023} (\bibinfo {year} {2016})},\ \Eprint
  {https://arxiv.org/abs/1609.01340} {arXiv:1609.01340 [gr-qc]} \BibitemShut
  {NoStop}%
\bibitem [{\citenamefont {Shipley}\ and\ \citenamefont
  {Dolan}(2016)}]{Shipley:2016omi}%
  \BibitemOpen
  \bibfield  {author} {\bibinfo {author} {\bibfnamefont {J.}~\bibnamefont
  {Shipley}}\ and\ \bibinfo {author} {\bibfnamefont {S.~R.}\ \bibnamefont
  {Dolan}},\ }\bibfield  {title} {\bibinfo {title} {{Binary black hole shadows,
  chaotic scattering and the Cantor set}},\ }\href
  {https://doi.org/10.1088/0264-9381/33/17/175001} {\bibfield  {journal}
  {\bibinfo  {journal} {Class. Quant. Grav.}\ }\textbf {\bibinfo {volume}
  {33}},\ \bibinfo {pages} {175001} (\bibinfo {year} {2016})},\ \Eprint
  {https://arxiv.org/abs/1603.04469} {arXiv:1603.04469 [gr-qc]} \BibitemShut
  {NoStop}%
\bibitem [{\citenamefont {Kostaros}\ and\ \citenamefont
  {Pappas}(2022)}]{Kostaros:2021usv}%
  \BibitemOpen
  \bibfield  {author} {\bibinfo {author} {\bibfnamefont {K.}~\bibnamefont
  {Kostaros}}\ and\ \bibinfo {author} {\bibfnamefont {G.}~\bibnamefont
  {Pappas}},\ }\bibfield  {title} {\bibinfo {title} {{Chaotic photon orbits and
  shadows of a non-Kerr object described by the Hartle\textendash{}Thorne
  spacetime}},\ }\href {https://doi.org/10.1088/1361-6382/ac7028} {\bibfield
  {journal} {\bibinfo  {journal} {Class. Quant. Grav.}\ }\textbf {\bibinfo
  {volume} {39}},\ \bibinfo {pages} {134001} (\bibinfo {year} {2022})},\
  \Eprint {https://arxiv.org/abs/2111.09367} {arXiv:2111.09367 [gr-qc]}
  \BibitemShut {NoStop}%
\bibitem [{\citenamefont {Kostaros}\ \emph {et~al.}(2024)\citenamefont
  {Kostaros}, \citenamefont {Papadopoulos},\ and\ \citenamefont
  {Pappas}}]{Kostaros:2024vbn}%
  \BibitemOpen
  \bibfield  {author} {\bibinfo {author} {\bibfnamefont {K.}~\bibnamefont
  {Kostaros}}, \bibinfo {author} {\bibfnamefont {P.}~\bibnamefont
  {Papadopoulos}},\ and\ \bibinfo {author} {\bibfnamefont {G.}~\bibnamefont
  {Pappas}},\ }\bibfield  {title} {\bibinfo {title} {{Fractal signatures of
  non-Kerr spacetimes in the shadow of light-ring bifurcations}},\ }\href
  {https://doi.org/10.1103/PhysRevD.110.024001} {\bibfield  {journal} {\bibinfo
   {journal} {Phys. Rev. D}\ }\textbf {\bibinfo {volume} {110}},\ \bibinfo
  {pages} {024001} (\bibinfo {year} {2024})},\ \Eprint
  {https://arxiv.org/abs/2405.09653} {arXiv:2405.09653 [gr-qc]} \BibitemShut
  {NoStop}%
\bibitem [{\citenamefont {{Kapec}}\ and\ \citenamefont
  {{Lupsasca}}(2020)}]{Kapec+2020}%
  \BibitemOpen
  \bibfield  {author} {\bibinfo {author} {\bibfnamefont {D.}~\bibnamefont
  {{Kapec}}}\ and\ \bibinfo {author} {\bibfnamefont {A.}~\bibnamefont
  {{Lupsasca}}},\ }\bibfield  {title} {\bibinfo {title} {{Particle motion near
  high-spin black holes}},\ }\href {https://doi.org/10.1088/1361-6382/ab519e}
  {\bibfield  {journal} {\bibinfo  {journal} {cqg}\ }\textbf {\bibinfo {volume}
  {37}},\ \bibinfo {eid} {015006} (\bibinfo {year} {2020})},\ \Eprint
  {https://arxiv.org/abs/1905.11406} {arXiv:1905.11406 [hep-th]} \BibitemShut
  {NoStop}%
\bibitem [{\citenamefont {{Gebhardt}}\ \emph {et~al.}(2011)\citenamefont
  {{Gebhardt}}, \citenamefont {{Adams}}, \citenamefont {{Richstone}},
  \citenamefont {{Lauer}}, \citenamefont {{Faber}}, \citenamefont
  {{G{\"u}ltekin}}, \citenamefont {{Murphy}},\ and\ \citenamefont
  {{Tremaine}}}]{Gebhardt+2011}%
  \BibitemOpen
  \bibfield  {author} {\bibinfo {author} {\bibfnamefont {K.}~\bibnamefont
  {{Gebhardt}}}, \bibinfo {author} {\bibfnamefont {J.}~\bibnamefont {{Adams}}},
  \bibinfo {author} {\bibfnamefont {D.}~\bibnamefont {{Richstone}}}, \bibinfo
  {author} {\bibfnamefont {T.~R.}\ \bibnamefont {{Lauer}}}, \bibinfo {author}
  {\bibfnamefont {S.~M.}\ \bibnamefont {{Faber}}}, \bibinfo {author}
  {\bibfnamefont {K.}~\bibnamefont {{G{\"u}ltekin}}}, \bibinfo {author}
  {\bibfnamefont {J.}~\bibnamefont {{Murphy}}},\ and\ \bibinfo {author}
  {\bibfnamefont {S.}~\bibnamefont {{Tremaine}}},\ }\bibfield  {title}
  {\bibinfo {title} {{The Black Hole Mass in M87 from Gemini/NIFS Adaptive
  Optics Observations}},\ }\href {https://doi.org/10.1088/0004-637X/729/2/119}
  {\bibfield  {journal} {\bibinfo  {journal} {\apj}\ }\textbf {\bibinfo
  {volume} {729}},\ \bibinfo {eid} {119} (\bibinfo {year} {2011})},\ \Eprint
  {https://arxiv.org/abs/1101.1954} {arXiv:1101.1954 [astro-ph.CO]}
  \BibitemShut {NoStop}%
\bibitem [{\citenamefont {{GRAVITY Collaboration}}\ \emph
  {et~al.}(2018)\citenamefont {{GRAVITY Collaboration}} \emph
  {et~al.}}]{Gravity+2018}%
  \BibitemOpen
  \bibfield  {author} {\bibinfo {author} {\bibnamefont {{GRAVITY
  Collaboration}}} \emph {et~al.},\ }\bibfield  {title} {\bibinfo {title}
  {{Detection of orbital motions near the last stable circular orbit of the
  massive black hole SgrA*}},\ }\href
  {https://doi.org/10.1051/0004-6361/201834294} {\bibfield  {journal} {\bibinfo
   {journal} {aap}\ }\textbf {\bibinfo {volume} {618}},\ \bibinfo {eid} {L10}
  (\bibinfo {year} {2018})},\ \Eprint {https://arxiv.org/abs/1810.12641}
  {arXiv:1810.12641 [astro-ph.GA]} \BibitemShut {NoStop}%
\bibitem [{\citenamefont {{Walker}}\ and\ \citenamefont
  {{others}}(2018)}]{Walker+2018}%
  \BibitemOpen
  \bibfield  {author} {\bibinfo {author} {\bibfnamefont {R.~C.}\ \bibnamefont
  {{Walker}}}\ and\ \bibinfo {author} {\bibnamefont {{others}}},\ }\bibfield
  {title} {\bibinfo {title} {{The Structure and Dynamics of the Subparsec Jet
  in M87 Based on 50 VLBA Observations over 17 Years at 43 GHz}},\ }\href
  {https://doi.org/10.3847/1538-4357/aaafcc} {\bibfield  {journal} {\bibinfo
  {journal} {\apj}\ }\textbf {\bibinfo {volume} {855}},\ \bibinfo {eid} {128}
  (\bibinfo {year} {2018})},\ \Eprint {https://arxiv.org/abs/1802.06166}
  {arXiv:1802.06166 [astro-ph.HE]} \BibitemShut {NoStop}%
\bibitem [{\citenamefont {{Breton}}\ \emph {et~al.}(2008)\citenamefont
  {{Breton}}, \citenamefont {{Kaspi}}, \citenamefont {{Kramer}}, \citenamefont
  {{McLaughlin}}, \citenamefont {{Lyutikov}}, \citenamefont {{Ransom}},
  \citenamefont {{Stairs}}, \citenamefont {{Ferdman}}, \citenamefont
  {{Camilo}},\ and\ \citenamefont {{Possenti}}}]{Breton+2008}%
  \BibitemOpen
  \bibfield  {author} {\bibinfo {author} {\bibfnamefont {R.~P.}\ \bibnamefont
  {{Breton}}}, \bibinfo {author} {\bibfnamefont {V.~M.}\ \bibnamefont
  {{Kaspi}}}, \bibinfo {author} {\bibfnamefont {M.}~\bibnamefont {{Kramer}}},
  \bibinfo {author} {\bibfnamefont {M.~A.}\ \bibnamefont {{McLaughlin}}},
  \bibinfo {author} {\bibfnamefont {M.}~\bibnamefont {{Lyutikov}}}, \bibinfo
  {author} {\bibfnamefont {S.~M.}\ \bibnamefont {{Ransom}}}, \bibinfo {author}
  {\bibfnamefont {I.~H.}\ \bibnamefont {{Stairs}}}, \bibinfo {author}
  {\bibfnamefont {R.~D.}\ \bibnamefont {{Ferdman}}}, \bibinfo {author}
  {\bibfnamefont {F.}~\bibnamefont {{Camilo}}},\ and\ \bibinfo {author}
  {\bibfnamefont {A.}~\bibnamefont {{Possenti}}},\ }\bibfield  {title}
  {\bibinfo {title} {{Relativistic Spin Precession in the Double Pulsar}},\
  }\href {https://doi.org/10.1126/science.1159295} {\bibfield  {journal}
  {\bibinfo  {journal} {Science}\ }\textbf {\bibinfo {volume} {321}},\ \bibinfo
  {pages} {104} (\bibinfo {year} {2008})},\ \Eprint
  {https://arxiv.org/abs/0807.2644} {arXiv:0807.2644 [astro-ph]} \BibitemShut
  {NoStop}%
\end{thebibliography}%
	
\end{document}